\documentclass{aa}  
\usepackage{natbib}
\usepackage{longtable}
\usepackage{array}
\usepackage{tablefootnote}
\usepackage{graphicx}
\usepackage{txfonts}
\usepackage[colorlinks=true,allcolors=blue]{hyperref}

\begin{document} 

\title{Neighbors, Not Kin: Kinematic evidence for tidal tails from NGC~7492 along the Sagittarius stream\thanks{Based on observations collected at the European Southern Observatory under ESO programme 111.253P.001.}}

   \author{Camila Navarrete
          \inst{1,2},\thanks{\email{camila.navarrete@oca.eu}},
          \and
          Álvaro Rojas-Arriagada\inst{3,4}
         \and
          Andrés E. Piatti\inst{5,6}
          \and 
          Julio A. Carballo-Bello\inst{7} \and
          Luca Sbordone\inst{8} \and 
          Richa Kundu\inst{9,10} \and 
          Vasily Belokurov \inst{11} \and 
          Sergey E. Koposov \inst{12, 11} \and
          Eduardo Vitral\thanks{Royal Society Newton International Fellow}\inst{12} \and 
          Pierre Boldrini\inst{13} \and 
          Pedro A. Palicio\inst{1}}
    
    \authorrunning{Navarrete et al. }

   \institute{Universit{\'e} C{\^o}te d'Azur, Observatoire de la C{\^o}te d'Azur, CNRS, Laboratoire Lagrange, Bd de l'Observatoire, CS 34229, 06304, Nice Cedex 4, France
  \and
  Centro de Investigación en Ciencias del Espacio y Física Teórica, Universidad Central de Chile, La Serena 1710164, Chile
  \and 
  Departamento de Física, Universidad de Santiago de Chile, Av. Victor Jara 3659, Santiago, Chile \and 
  Center for Interdisciplinary Research in Astrophysics and Space Exploration (CIRAS), Universidad de Santiago de Chile, Santiago, Chile \and 
  Instituto Interdisciplinario de Ciencias B\'asicas (ICB), CONICET-UNCuyo, Padre J. Contreras 1300, M5502JMA, Mendoza, Argentina \and 
  Consejo Nacional de Investigaciones Cient\'{\i}ficas y T\'ecnicas, Godoy Cruz, 2290, C1425FQB,  Buenos Aires, Argentina \and
  Instituto de Alta Investigación, Universidad de Tarapacá, Casilla 7D, Arica, Chile \and 
  ESO–European Southern Observatory, Alonso de C\'ordova 3107, Vitacura, Santiago, Chile \and 
  Miranda House, University of Delhi, Delhi-110007, India \and 
  Inter University Centre for Astronomy and Astrophysics, Pune, Maharashtra, India \and 
  Institute of Astronomy, University of Cambridge, Madingley Road, Cambridge CB3 0HA, UK \and  
  Institute for Astronomy, University of Edinburgh, Royal Observatory, Blackford Hill, Edinburgh EH9 3HJ, UK \and
  LIRA, Observatoire de Paris, Université PSL, Sorbonne Université, Université Paris Cité, CY Cergy Paris Université, CNRS, 92190 Meudon, France} 

  \titlerunning{Tidal tails from NGC 7492 along the Sgr stream}
 \authorrunning{C. Navarrete et al.}

   \date{Received ... ..., 2025; accepted ... ..., 2026}
  \abstract
   {The formation, extension, and morphology of extra-tidal stars around globular clusters depend on the internal kinematics of the host cluster and the influence of the Galactic potential. Tracing the kinematics of faint tidal tails sheds light on their formation and contribution to the Milky Way halo assembly.}
   {NGC 7492 is an outer halo globular cluster with conflicting evidence regarding the presence of tidal tails. If present, the tails are expected to be faint and embedded in the Sagittarius tidal stream, located at a similar heliocentric distance but having different kinematics.}
   {We carried out a GIRAFFE spectroscopic follow-up of ten fields at the expected location of tidal tails associated with NGC 7492 to obtain the radial velocity component. \textit{Gaia} parallaxes were used to remove foreground contaminants, while only loose constraints on \textit{Gaia} proper motion were applied to select the targets. From the high-resolution spectra, radial velocities and metallicities were derived for more than 700 stars, from the red giant branch down to the upper main sequence.} 
   {Cluster and extra-tidal stars were identified based on their proper motions, radial velocities, and metallicities. This population extends over at least $\sim$1.8 deg from the cluster center, confirming the tidal tails previously detected only photometrically. The extra-tidal stars are located at positions consistent with spray-particle models for the cluster's disruption. The Sagittarius stream is also clearly identified through its distinct proper motion distribution, radial velocities and more metal-rich population.}
   {Despite the low spatial density of extra-tidal stars, the kinematic signature of the cluster is clearly detected, confirming the presence of tidal tails overlapping on the sky with the Sagittarius stream, though the two structures are physically unrelated. The present spectroscopic dataset provides a robust basis for future studies aimed at extending the characterization of the tails, both in extension and limiting magnitude.}

   \keywords{stars: kinematics and dynamics - Galaxy: halo - globular clusters: individual: NGC 7492 - Galaxy: structure}

   \maketitle
%
 \defcitealias{Navarrete17}{N17}

\section{Introduction}

The disruption of globular clusters (GCs) by the tidal forces exerted by the Milky Way (MW) produces extended tidal tails that often trace the clusters’ orbit in the form of thin, elongated, S-shaped streams. These structures serve as valuable dynamical tracers, offering insight into both the internal evolution of clusters and the external tidal field in which they evolve \citep[e.g.][]{Montuori07, Hozumi15, Kupper15, Balbinot18}. 

N-body simulations show that tidal tails form primarily due to repeated interactions with the densest Galactic components, such as the bulge and disk, and can develop into multi-component features after successive pericentric passages \citep{Montuori07, Hozumi15}. The extension, width and velocity dispersion of tidal tails provide useful insights into the properties of the progenitor, such as its mass, metallicity distribution and origin. For instance, based on N-body simulations, \cite{Malhan21, Malhan22} found that tidal tails from accreted GCs tend to be thicker and kinematically hotter than those from GCs born in situ, and there are significant differences between tidal tails originating from accreted GCs inside cuspy or cored dark-matter density profiles of their host dwarf galaxies. 

Wide-field photometric surveys have played a central role in revealing tidal features around GCs. The tidal tails of Palomar 5 are a well-studied example \citep{Odenkirchen01, Odenkirchen03, Ibata16, Bonaca20}, while hints of extra-tidal substructures have been reported in other systems \citep[see the compilation in][]{PiattiCarballo20}. The advent of the \textit{Gaia} mission \citep{Gaia2016, Gaia2018} has enabled a dramatic improvement in the separation of overlapping stellar populations, particularly in GCs located in highly dense regions, uncovering a wide variety of low surface-brightness features associated with GCs \citep{Carballo19, Sollima20, Kundu21, Yang22}. Despite these advances, the formation of tidal tails appears to be governed by a complex combination of internal cluster properties, including mass, structural parameters, and internal dynamical evolution driven by two-body relaxation \citep{Meylan97, Baumgardt03, Weatherford23}, and external effects related to the Galactic tidal field, disk and bulge shocks, and cluster orbital properties \citep{Gnedin97, Kupper10}. A recent analysis of literature data by \citet{Zhang22} found no strong correlation between the presence of tidal tails and kinematical and structural parameters from the parent GCs, highlighting the need for case-by-case dynamical modeling to understand their origins \citep[see as well][]{PiattiCarballo20}.

NGC 7492 is a metal-poor GC \citep[{[}Fe/H{]} $\sim$ –1.82 dex;][]{Cohen05} located in the outer halo of the Galaxy at a heliocentric distance of $\sim$25 kpc \citep{FigueraJaimes13}, and projected against the Sagittarius (Sgr) stellar stream, a halo substructure generated by the disruption of the Sgr dwarf spheroidal galaxy \citep[][]{Ibata01, Majewski03, Belokurov06, Koposov12, Belokurov14, Navarrete17_sgr, Antoja20, Ramos22}. Although extended populations consistent with the Sgr stream have been detected near the cluster in photometric data \citep{Julio14, Sollima18}, these stars are significantly more metal-rich \citep[–1.5 $<$ {[}Fe/H{]} $<$ –0.5;][]{Hayes20}, and dynamical analyses have argued against a physical association between NGC 7492 and the Sgr galaxy \citep{Julio18, Arakelyan20, Penarubia21}. It is considered an accreted GC, most likely associated with the Helmi stream \citep{Callingham22} rather than to be associated with the Sgr galaxy or being an in-situ GC of the MW. 

The first detection of tidal tails around NGC 7492 was reported by \citet[][hereinafter \citetalias{Navarrete17}]{Navarrete17}, who used a modified matched-filter technique combining color-magnitude diagram (CMD) information across multiple bands. They identified tails extending $\sim$3.5 degrees on the sky (or $\sim$1.5 kpc), distinguishing the GC population from the overlapping Sgr stream. However, at the time, the lack of proper motion data limited the ability to cleanly separate cluster stars from field contamination. Follow-up radial velocity (RV) measurements by \citet{Julio18} identified at least three velocity components in two small fields outside the cluster's nominal tidal radius \citep[r$_{\rm t}$ = 9.2 arcmin,][]{CarballoBello12} corresponding to the cluster, the Sgr stream, and the MW field population (see their Fig. 4). More recent photometric studies have yielded mixed results. \citet{Munoz18} and \citet{Zhang22} did not recover a prominent cluster population within 30 arcminutes of the cluster. Based on deep photometry from the DELVE survey \citep{delve22}, \cite{Chiti25} recovered a photometric excess beyond the cluster Jacobi radius\footnote{The Jacobi radius $r_J$ is defined as the distance from the center of the cluster to the first Lagrangian point. Stars located beyond the tidal radius but inside the Jacobi radius are still bound to the cluster.}, in the same direction as the tails recovered by \citetalias{Navarrete17}. Based on \textit{Gaia} DR3 proper motions, \citet{Ibata24} identified 26 extra-tidal stars forming a thin, stream-like structure spatially associated to NGC 7492, although only one star had an RV measurement. Similar results were recently reported in \cite{Chenstreams} were more than 90 stars inside and outside the tidal radius of the cluster were recovered based on \textit{Gaia} astrometry and photometry and in \cite{Wang26} where DESI Legacy Survey \citep{Dey2019} wide-field photometry and matched-filtering techniques were used to recover its tidal tails. 

With the availability of high-precision astrometry from \textit{Gaia} DR3 \citep{Gaia2016, Gaia2018, Gaia2023}, it is now possible to efficiently separate the overlapping stellar populations around GCs using proper motion data alone \citep[e.g.,][]{Kundu22}. However, to fully disentangle the cluster, field, and Sgr stream populations, and to confirm the dynamical status of the tails, spectroscopic observations are essential to obtain full 6D phase-space information. In particular, RVs and metallicities are needed to complement astrometric selections and construct detailed kinematic profiles of potential tidal debris. Moreover, as stellar streams from accreted GCs are expected to be kinematically hotter and wider on the sky than those from in situ GCs \citep{Malhan21, Malhan22}, a statistically significant sample, unbiased in the kinematic selection, is needed to recover any potential tidal tail in this cluster successfully.

In this work, we present an extensive spectroscopic campaign along the expected location of the tidal tails of NGC 7492 using the GIRAFFE@FLAMES spectrograph, targeting stars in the cluster's outskirts. By combining these data with \textit{Gaia} DR3 astrometry and photometry, we aim to (i) isolate the cluster population from the surrounding Sgr and field stars, (ii) derive RV and metallicity measurements for a large sample of candidates, and (iii) compare the resulting velocity maps with predictions from numerical simulations for the disruption of this cluster. In this way, we attempt to confirm or reject, based on a kinematical analysis, the existence of the tidal tails originally identified in \citetalias{Navarrete17}. This paper is organized as follows. Section~\ref{sec:data} describes the target selection and the GIRAFFE@FLAMES data acquisition and reduction. Section~\ref{spec_analysis} presents the RV and metallicity measurements and the comparison with literature values. The separation of stars likely associated to the cluster and those most likely related to the Sgr stream, as well as the spatial distribution, CMD, proper motions, RVs and metallicities of the extra-tidal star candidates are presented in Section~\ref{results}. In Section~\ref{discussion}, we compared the distribution of the extra-tidal star candidates with numerical models for the disruption of the cluster and its orbit. We conclude by summarizing our results and discussing prospects of this work in Section~\ref{summary}. The individual RVs and metallicity measurements for the stars associated with the cluster and the Sgr stream are made public as part of this work, while the complete sample (including MW field stars) will be available from the authors upon reasonable request .

\section{Data}\label{sec:data}

\subsection{Observations}

\begin{figure}
\includegraphics[width=0.48\textwidth]{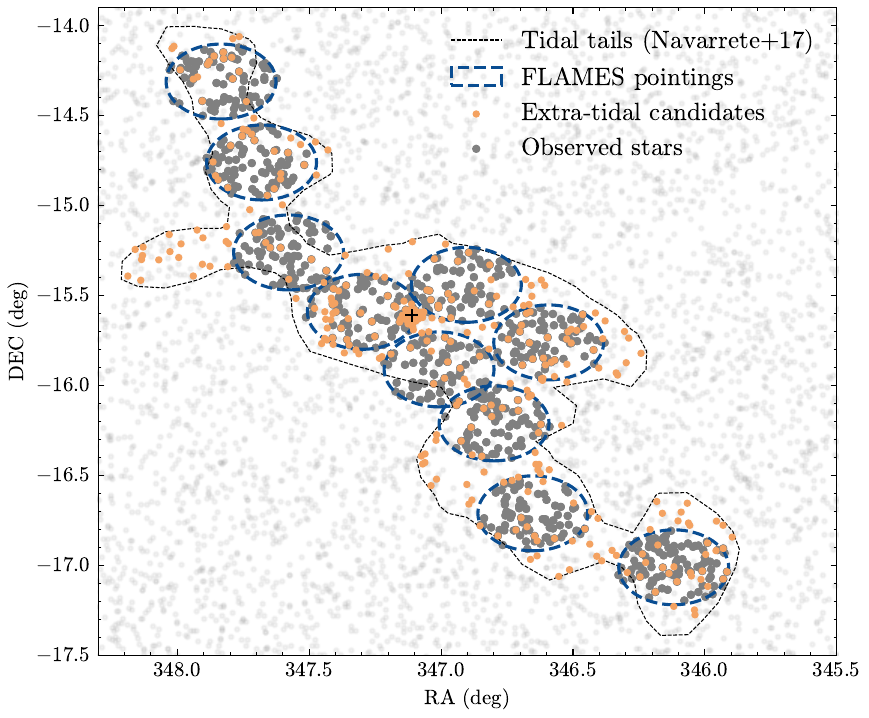}
\caption{Spatial distribution of \textit{Gaia} DR3 sources following the trace of the tidal tails from \citetalias{Navarrete17} (dashed grey contour). The center of the cluster is shown with a black plus symbol. Each of the FLAMES pointings is marked as dashed blue circles, while the preliminary extra-tidal candidates, based on the CMD and isochrone of the cluster, are shown as orange points. The observed targets are shown as grey dots.}\label{fig:spatial}
\end{figure}

To trace the potential tidal tails emerging from NGC 7492, we followed the contours outlined in \citetalias{Navarrete17} and selected several fields along the tails to obtain spectra for as many stars as possible in each field. The ten fields shown in Fig.~\ref{fig:spatial} were observed with the ESO GIRAFFE@FLAMES spectrograph \citep{Pasquini02}, mounted on UT2 at the Very Large Telescope (VLT, Cerro Paranal, Chile), having a diameter of 25 arcmin each. The location of each field was decided in order to cover the full extension of the tails, while avoiding the central parts of the cluster. In Fig~\ref{fig:spatial}, an initial selection of extra-tidal candidates, based on \textit{Gaia} proper motions and CMD distribution, is shown as orange circles, to demonstrate the low stellar density per field of the most promising candidates identified from \textit{Gaia} \citepalias[with original contours being those derived based on Pan-STARSS photometry in][]{Navarrete17}. The selection of the observed targets, as dark grey circles, is explained below.

Since stars belonging to cluster tidal tails or extra-tidal structures are in the process of escaping their clusters, their velocity behavior could be somewhat different from that of stars still bound to the cluster \citep{Boldrini21, Wang21}. Therefore, we did not perform any selection in proper motion, aiming to do a blind map of each field, allowing Sgr stream stars or any field star to be observed as well. The only selection criteria applied was to remove stars with exceedingly high transverse motion and large parallaxes through the exclusion of sources with \textit{Gaia} DR3 max($|\mu_{\alpha} \cos{\delta}; \mu_{\delta}|) \geq 10$ mas yr$^{-1}$ and parallax $>$ 0.5 mas. The limiting magnitude to allocate a source into a fibre was set to G = 21.0 mag.

Even with this relaxed cut in proper motion and no constraint on the CMD, the number of sources per field was generally below the maximum number of fibres available. Therefore, for the observations, we decided not to impose a selection in the CMD and observe instead all the sources available per field, including red sources and potential blue horizontal branch (BHB) stars. Table~\ref{tab:fields} presents the central coordinates of the ten fields, the number of fibres allocated to stars and the median S/N of all the fibres observed in each field. In each field, $\sim$10 fibers were allocated to sky positions to obtain a median sky spectrum. This sky spectrum, one per field, was used to remove the telluric emission lines present in the science data. 

The observations were carried out between May and July 2023, in Service Mode (run ID: 111.253P.001; P.I. Navarrete). The HR21 setup was used, with a resolution of $\sim$18\,000, since it is centred on the Ca II triplet (CaT, wavelength range from 8480 to $\sim$9000 {\AA}). The observations consisted of two exposures of 2760 s ($\sim$45 min) each, per field. The S/N ratio is magnitude dependent, ranging from $\approx$4 (\textit{Gaia} G band = 20.6 mag) to 190 (G = 12.7 mag). The reduced and extracted 1D spectra were retrieved from the GIRAFFE ESO Phase 3 Data Products\footnote{\url{https://doi.eso.org/10.18727/archive/27}.}. The raw data was reduced with the ESO pipeline \texttt{giraf-2.16.9}, which performs cosmic rays removal, bias subtraction, bad pixel masking, flat-field correction and wavelength calibration. The top panel in Fig.~\ref{fig:spectrum} shows the observed spectrum for a star with S/N = 12 and the medium sky spectrum from all the fibres in the same field. The reduction pipeline for GIRAFFE does not consider sky subtraction, although the contribution of sky lines may be significant, particularly for spectra with low to moderate S/N in which the flux of the skylines can be as strong as the stellar flux. 

\begin{table}
    \centering
    \caption{Observed fields with GIRAFFE. The field ID, equatorial coordinates of the center, the number of fibres allocated to stars and the median S/N of the co-added spectra are listed.}
    \begin{tabular}{lclcc}
   \hline
   ID  & R.A.      &  Dec.             & N fibres & Mean S/N \\
       & (J2000.0) & (J2000.0)         &          &  \\
   \hline
Field  1  & 23:04:28.37 & -17:00:41.3  & 89 & 15 \\
Field  2  & 23:06:22.11 & -15:45:41.3  & 72 & 19 \\
Field  3  & 23:06:36.54 & -16:42:41.3  & 72 & 18 \\
Field  4  & 23:07:11.82 & -16:12:41.3  & 77 & 25 \\
Field  5  & 23:07:37.00 & -15:26:30.0  & 76 & 17 \\
Field  6  & 23:08:01.84 & -15:54:41.3  & 73 & 16 \\
Field  7  & 23:09:11.65 & -15:35:29.3  & 81 & 38 \\
Field  8  & 23:10:18.75 & -15:15:41.3  & 73 & 23 \\
Field  9  & 23:10:43.31 & -14:45:41.3  & 79 & 16 \\
Field 10  & 23:11:20.18 & -14:18:41.3  & 74 & 21 \\
\hline
    \end{tabular}
    \label{tab:fields}
\end{table}

\subsection{Sky subtraction}

\begin{figure}
\includegraphics[width=0.5\textwidth]{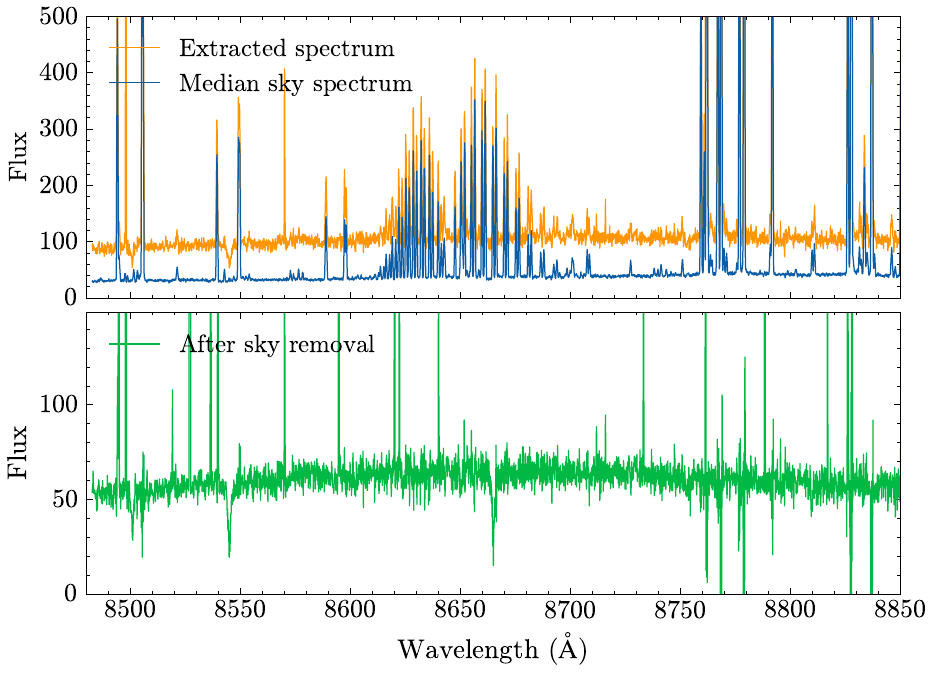}
\caption{Example of the sky subtraction. Top panel: The extracted spectrum for a star with S/N = 12 (Gmag = 19.3 mag, orange line) and the associated median sky spectrum for the field, from the median average of the individual sky spectra (blue line). Bottom panel: Co-added stellar spectrum after sky removal, with S/N = 17. Most of the sky lines are effectively removed, particularly the ones around 8650 {\AA}.}\label{fig:spectrum}
\end{figure}

To remove the skylines, a median sky spectrum was constructed from the individual sky fibres, for each field. A 3$\sigma$ clipping was performed in the individual sky spectrum to remove cosmic rays and spurious detections. This median sky spectrum was scaled, with respect to the observed spectrum, and then subtracted following the procedure outlined in \cite{Battaglia08}. The heliocentric correction in the wavelength of the science spectra was removed first, and after the sky subtraction, applied back.  Then, the two individual spectrum for each star were combined, first interpolating each of them in wavelength, and then obtaining a mean spectrum. In the case of the observations of the Field 5, the first exposure has a considerably lower S/N, under similar atmospheric conditions, than the second one, most likely due to an inaccurate centering of the fibres. In this case, only the second exposure was considered. Figure~\ref{fig:spectrum} shows an example of this procedure in which the extracted spectrum and the median sky spectrum (upper panel) are shown for a star having S/N=12. In the bottom panel, the spectrum after the sky subtraction and the co-adding with the second exposure is shown, reaching a S/N=17. Even though some residuals are still present (most likely cosmic rays), the majority of skylines around the CaT lines are removed. 

Combining the individual sky spectra has the caveat that any spatial variation in the sky emission is not taken into account. However, in order to avoid over-subtracting the sky emission, we preferred to perform a sky subtraction that leaves some isolated emission lines, which can be masked out before obtaining the RVs and metallicities.

\section{Spectroscopic analysis}\label{spec_analysis}

\subsection{Radial velocities}
Once the sky-subtracted and coadded spectra are ready, they were normalized using a spline fitting with iterative clipping. The RVs were then computed with a custom-made algorithm using a small grid of synthetic spectra as templates. They broadly cover the expected range of atmospheric parameters  and metallicity of the sample (T$_{\rm eff}$: 4300--6500 K, $\log(g)$: 1.3--4.6 dex and [Fe/H]: -1.5--+0.0 dex). For each spectrum, a first cross-correlation was performed with a randomly selected template to determine a preliminary RV value. The corrected observed spectrum was then compared to the whole grid, and the template with the lower $\chi^2$ was selected to compute the final refined velocity. When calculating the cross-correlation function (CCF), the algorithm refined the maximum value by fitting a Gaussian curve to the largest peak. In the case of hot stars (i.e., T$_{\rm eff} \geq$ 7000 K), we performed a cross-correlation with a different and larger grid covering T$_{\rm eff}$: 7000-15000 K, and rotational velocity levels up to 120 km s$^{-1}$, to optimize the RV estimate.

We estimated errors in RV by performing repeated estimates sampling in flux errors. For each spectrum, we produced 500 Gaussian resamplings around each flux value according to the scale set by the flux error. After the RVs were computed for the entire set of resampled spectra, using only the best $\chi^2$ template, the error associated with the original RV measurement was computed as the robust standard deviation of all sampled RVs. In order to avoid the effect of outliers, we computed the median absolute deviation (MAD) adopting the known relation $\sigma=1.4826\cdot MAD$. The comparison between our RV measurements and those available in the literature is presented in detail in Appendix~\ref{app:rv_comp}. We found that the derived RVs are consistent with literature values, within uncertainties.

\subsection{Atmospheric parameters and metallicity}
The atmospheric parameters, effective temperature $\mathrm{T_{eff}}$, surface gravity $\log(g)$ and metallicity [M/H] were estimated through full spectrum fitting against a grid of synthetic spectra. To this end, we used the code FERRE \citep{AllendePrieto2006}. FERRE compares observed fluxes against a grid of synthetic templates, allowing for interpolation, and finds the best match in a $\chi^2$ sense. To ensure consistency, both observations and models were normalized using polynomials (of fourth degree here). For this analysis, we compiled grids in three dimensions ($\mathrm{T_{eff}}$, $\log(g)$ and metallicity). The library of synthetic spectra was computed with Turbospectrum\footnote{Available at \url{https://github.com/bertrandplez/Turbospectrum2019}}, adopting MARCS "standard abundance" models\footnote{In these models, alpha-enhancement is set according to metallicity following the Galactic trend. The website of MARCS is \url{https://marcs.astro.uu.se/}} \citep{Gustafsson2008}, and atomic and molecular data from the Gaia-ESO Survey (GES) linelist \citep{Heiter2021}. Microturbulence was set according to the ($\mathrm{T_{eff}}$, $\log g$ and [M/H]) of the model, according to the GES relation. While compiling the grids, the synthetic fluxes were convolved to the resolution of GIRAFFE HR21 with a Gaussian kernel. As the grids must be regular (all combinations of parameters, given their regular samplings, should be available), we compiled three grids covering approximately the effective temperature ranges of F, G and K, and M stars. 

The observed spectra were interpolated to match the wavelength sampling of the grids and were fitted against the three grids. For each star, we selected the set of parameters corresponding to the fit with the lowest $\chi^2$. FERRE allows the inclusion of flux errors in the computations. They can be used to exclude flux points if high errors are assigned to them. We included flux errors and ran the analysis twice. In the first run, points identified as cosmic rays and other cosmetics were masked in the individual spectra. Once the best synthetic fits were obtained, the dispersion of the residuals for each pixel over the entire sample were computed, and the MAD was computed as a robust dispersion estimation. Using these values, we identified three regions of larger systematic discrepancies between models and observed spectra. They amount to 6\% of the flux points, and are probably due to insufficient molecule modelling from the linelists included in our syntheses. A second run of the analysis was run masking these regions.

\begin{figure}
\includegraphics[width=0.45\textwidth]{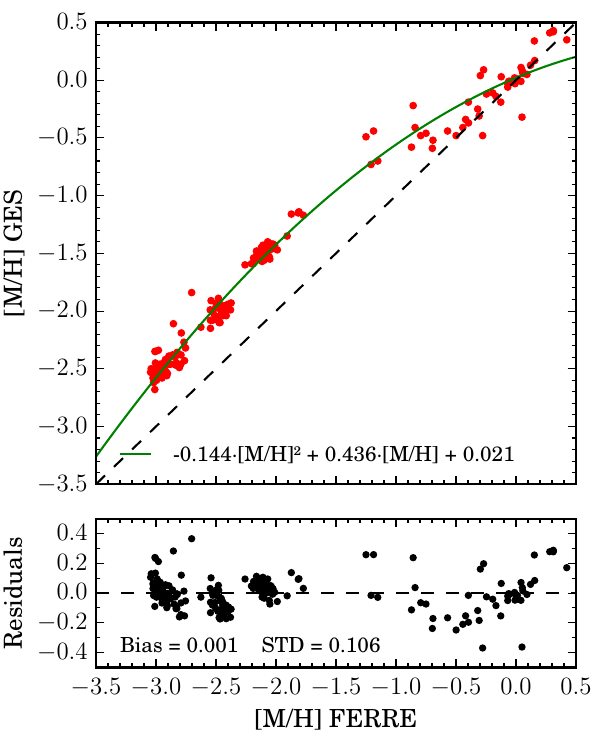}
\caption{Calibration of the metallicity scale. Top panel: comparison of metallicities for the calibration sample. The green solid line represents the best quadratic fit to the data, whose coefficients are displayed below. Bottom panel: residuals of the fit.}\label{fig:ferre_vs_ges}
\end{figure}

We calibrated our metallicities to the astrophysical scale of the GES. To this end, we retrieved a sample of 216 stars in the field of the clusters M~12, M~15, NGC~4833 and NGC 6752. We compared our metallicities with respect to those provided in the Public Data Release 5.1 of the GES, and derived a quadratic relation to bring our results to their astrophysical scale. We note that in every cluster, a small fraction of foreground stars allowed us to have a sparse sampling up to supersolar metallicities. In Fig.~\ref{fig:ferre_vs_ges}, we compare the metallicities of the GES and ours for the calibration sample. The calibrating quadratic polynomial appropriately captured the relation between both sets of estimates over the entire metallicity range. The residuals are flat, the bias is null and the dispersion is within the expected typical uncertainties of 0.1~dex. 

\begin{figure}
\includegraphics[width=0.5\textwidth]{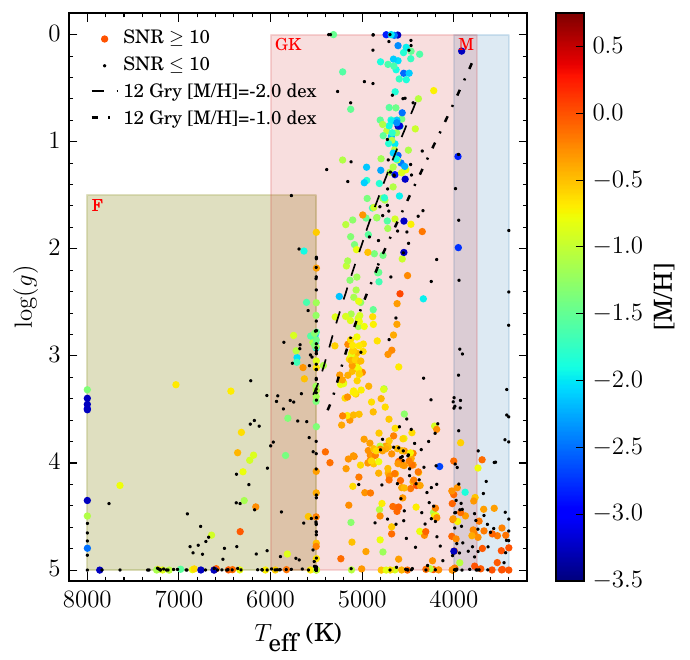}
\caption{Effective temperature versus surface gravity, as estimated from full spectrum fitting. Points corresponding to results derived from spectra with SNR larger than 10 are color coded by calibrated metallicity. As a reference, PARSEC isochrones \citep{Bressan12} of 12 Gyr with [M/H]=--1.0  and --2.0 dex are displayed with black lines. The colored areas show the location of the three grids used to analyze the spectra.}\label{fig:hrd}
\end{figure}

Figure~\ref{fig:hrd} shows the $\mathrm{T_{eff}}$ vs $\log(g)$ diagram for our entire sample. Only results for stars with spectra with SNR$\geq10$ are color coded by calibrated metallicity. This threshold is defined based on the large [M/H] error obtained for spectra having lower SNR, see Appendix~\ref{app:snr}. The range covered by the three synthetic spectral grids are depicted with colored areas. Hot stars, such as BHB stars, were recovered at the edge of the grid, producing a vertical distribution at $\mathrm{T_{eff}} \approx$ 8000 K. Due to this, we could not find a good parametrization for these spectra and, therefore, we consider only their derived RVs for the rest of the analysis. From the Kiel diagram, it can be seen that most of the metal-rich stars have large surface gravities, consistent with being dwarf stars, in contrast with most of the metal-poor sources. This metal-rich population is most likely dwarf stars from the Galactic foreground, not removed by the \textit{Gaia} parallax cut. Their RVs are consistent as well with being dwarf stars associated with the local disk.

\section{Results}\label{results}

\subsection{The kinematic component: tidal tails and the Sgr stream}\label{sec:sgr}

NGC 7492 is known to be immersed in the Sgr stream, at the same heliocentric distance \citep{CarballoBello14, Sollima18}, although not associated with it \citep{Julio18}. Based on numerical models, the Sgr stream stars should have a distinct velocity signature at the position of the cluster. In particular, \cite{Julio18} found the signature of two velocity components, based on 71 stars observed in two fields beyond the tidal radius of the cluster, potentially associated with the Sgr stream. The two peaks found were at heliocentric RVs $\sim -$110 km s$^{-1}$ and 125 km s$^{-1}$, coincident with the predictions from the numerical model of \cite{LM10} for the Sgr's trailing arm and the leading arm, respectively. The signature of the leading arm at $\sim$125 km s$^{-1}$ is also predicted by the model of \cite{Penarrubia10}. As we did not perform any exclusion based on proper motion information, the velocity distribution of the ten fields is expected to include Sgr stream stars from both the trailing and leading arms. 

\begin{figure}
\includegraphics[width=0.48\textwidth]{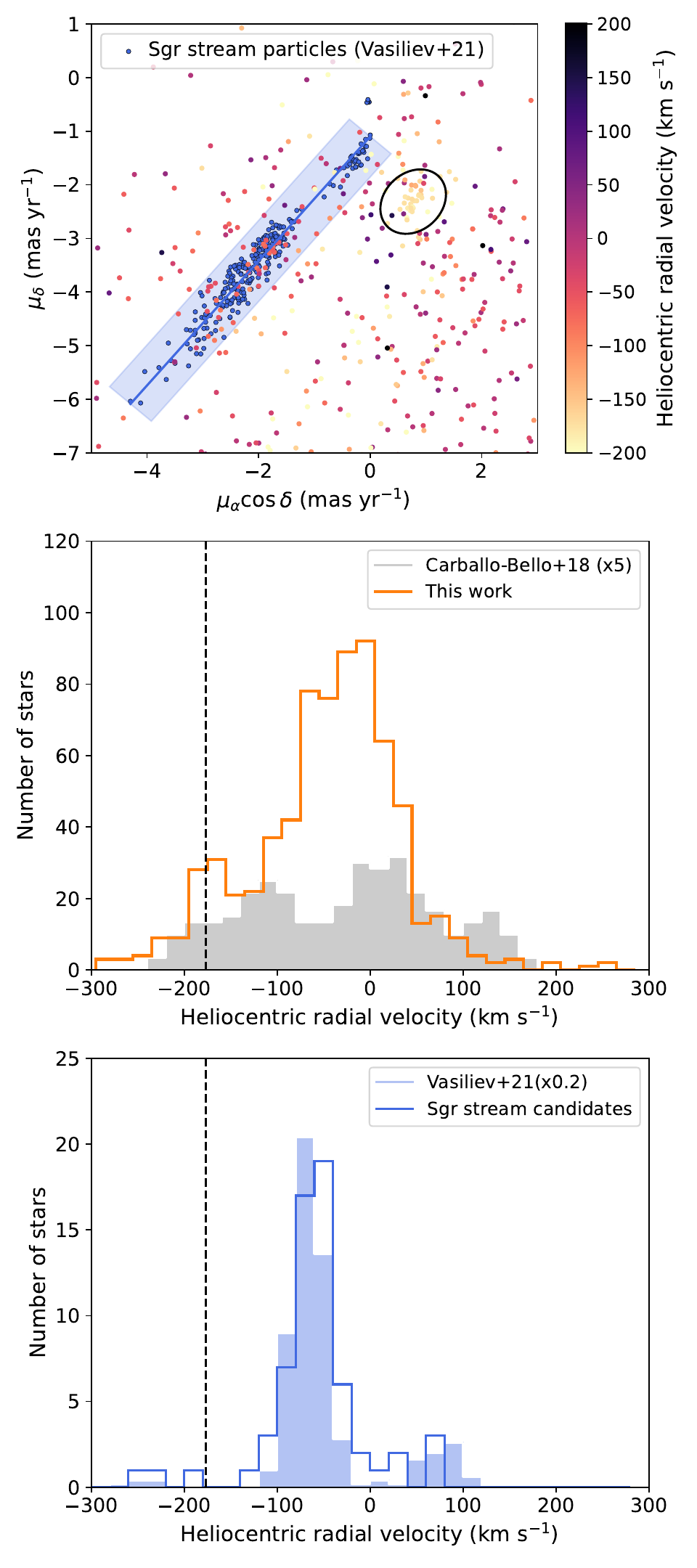}
\caption{RV distribution of our sample. Top panel: \textit{Gaia} DR3 proper motions of our stars (colored based on the heliocentric RV), and for Sgr stream particles \citep[][blue dots]{Vasiliev21} within 5 deg from the cluster center. The median proper motion of the cluster and the Sgr component are marked with a black ellipse and a blue solid line, respectively. Middle panel: RV distribution (orange line) for the observed stars. The observed RVs of stars outside the tidal radius, from \citet{Julio18}, are shown as a grey histogram (multiplied by five to enhance the visibility). The mean RV of the cluster is shown as a vertical dashed black line. Bottom panel: The predicted RVs from \cite{Vasiliev21} for particles within 5 deg from the cluster are shown as a blue histogram (divided by five for visibility), while the RVs of the stars with Sgr-like proper motions are shown as the blue, unfilled histogram.}\label{fig:tail_stream}
\end{figure}

The top panel of Figure~\ref{fig:tail_stream} shows the proper motion distribution of the observed stars, colored based on their heliocentric RV.  The signature of the cluster, centered at ($\mu_{\alpha}\cos{\delta}$, $\mu_{\delta}$) = (0.7, $-$2.3) mas yr$^{-1}$ \citep[black ellipse\footnote{The center, semi-major and semi-minor axes of the proper motion distribution for all the stars inside the cluster's tidal radius in \textit{Gaia} DR3 (i.e., not only restricted to our targets) were derived as in \cite{Kundu22}, obtaining ($\mu_{\alpha}\cos{\delta}$, $\mu_{\delta}$) = (0.77$\pm$0.18, $-$2.31$\pm$0.22) mas yr$^{-1}$.},][]{VasilievBaumgardt21}, has the expected mean velocity for the cluster stars, at $\sim-$180 km s$^{-1}$ \citep{BaumgardtHilker18}. The proper motion distribution for the Sgr stream particles from \cite{Vasiliev21}, located up to 5 deg from the cluster, is shown as blue points. The Sgr stream particles define a clear trend, different from the motion of the stars associated with the cluster. To isolate the stars most likely associated with the Sgr stream, we fitted a straight line to the proper motion distribution of these Sgr's particles, shown as a blue solid line in the top panel of Fig.~\ref{fig:tail_stream}, and defined a range of $\pm$3$\sigma$ (shaded blue area), being $\sigma$ the standard deviation of the residuals of the fit. 

The RV distribution of the observed stars in this work, having proper motions inside the Sgr's region, is shown in the bottom panel of Fig.~\ref{fig:tail_stream} (blue line histogram) and the predicted RVs from the simulation of \cite{Vasiliev21} in the same region (filled blue histogram, divided by a factor of five for better visibility). There is a remarkable agreement between the predicted and observed RVs, recovering the signature of the trailing arm and a handful of stars possibly associated with the leading arm. Not only do the peaks of the two distributions seem to agree, but the ratio between both components, being the trailing arm significantly more abundant than the leading, is well recovered based on the observed proper motions. A few stars have large RVs, that could be associated with a second trailing arm of the Sgr stream, or high RV halo stars. There is one star, \textit{Gaia} DR3 source ID 2409573650597520512, that has a RV = --188.5 km s$^{-1}$ similar to the cluster population (the bin close to the vertical dashed line). In the next section, when selecting stars associated to the cluster, this star is recovered as well as it has consistent proper motions, RV and position in the CMD. Therefore, we exclude it as a potential Sgr's stream star.

In the middle panel of Fig.~\ref{fig:tail_stream}, the RV distribution of all the observed stars is shown with an orange histogram. The majority of the sample has RVs centred on 0 km s$^{-1}$, being most likely MW stars. There is a second peak at the expected RV of the cluster (vertical dashed black line). The RV distribution of the two fields analysed in \cite{Julio18} is shown as a grey histogram. Following that work, this histogram was constructed using a bin size of 20 km s$^{-1}$ and smoothed with a boxcar average with a width of three bins. It was multiplied by a factor of five for better visibility. The boxcar was not applied to the other histograms. 

\begin{figure}
\includegraphics[width=0.5\textwidth]{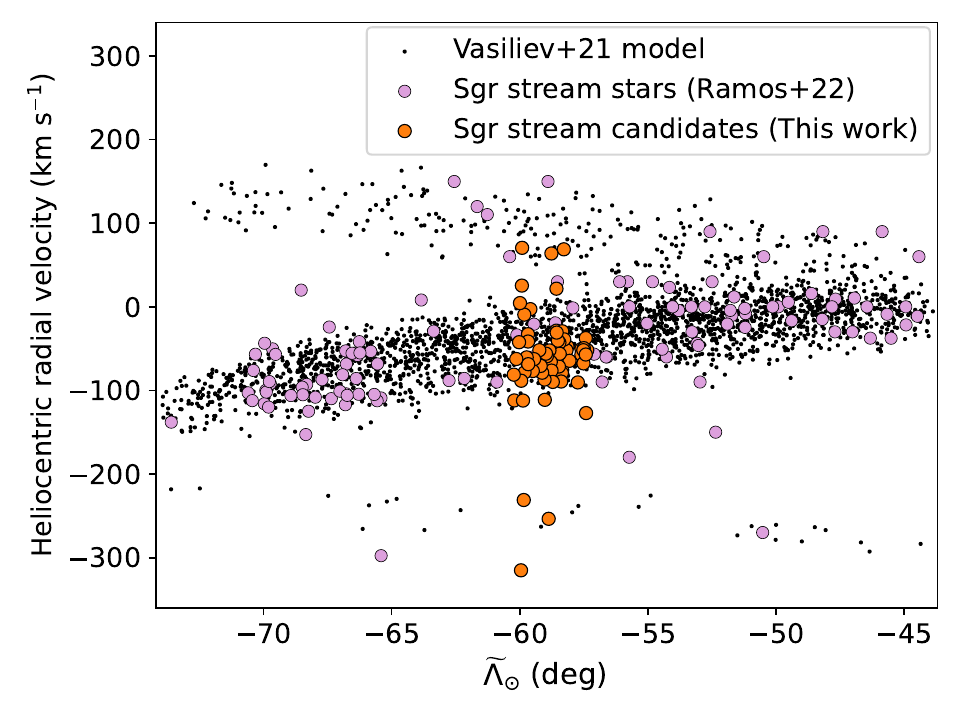}
\caption{Heliocentric RVs for the simulated Sgr particles from \citet{Vasiliev21} located up to 15 degrees from the cluster center (black dots) as a function of the Sgr stream longitude $\widetilde{\Lambda}_{\odot}$. The trailing arm is the most prominent branch of the Sgr stream at this position, while the model predicts as well an older leading arm (at RV $\sim$100 km s$^{-1}$) and trailing arm (RV $\sim-$250 km s$^{-1}$). Sgr stream stars up to 15 deg from the cluster from \citet{Ramos22} are shown as pink circles while the selected Sgr stream candidates are shown in orange circles.}\label{fig:lambda_vgsr}
\end{figure}

In Figure~\ref{fig:lambda_vgsr}, the heliocentric RV of the Sgr stream candidates is shown as a function of the Sgr stream longitude $\widetilde{\Lambda}_{\odot}$ \citep[as defined in][]{Vasiliev21}. The stream candidates from \textit{Gaia} eDR3 isolated in \cite{Ramos22}, located up to 15 deg from the cluster center, and the Sgr stream particles from the model of \cite{Vasiliev21}, are shown as pink and black dots, respectively. Most of the Sgr stream candidates among our observed stars have RVs consistent with the trailing arm, which is the most populated branch of the stream at this position of the sky. Only a few have velocities consistent with the leading arm, and the two stars at RV $\sim -200$ km s$^{-1}$ could be tentatively associated with the older trailing arm predicted by the simulations. Performing a cross-match between our targets and the Sgr member candidates from \cite{Ramos22}, we identified 28 stars in common, none of which have reported RVs in the literature. They have a high probability of being associated with the faint branch of the Sgr arm, according to the probability estimated in \cite{Ramos22} (\texttt{ProbFaint} $\geq$ 0.7). The properties of these Sgr stream candidate observed in this work, and those selected based on proper motion, are presented in the Appendix~\ref{app:tbc}.

\subsection{The tidal tails of NGC 7492}

From the RV distribution shown in the middle panel of Fig.~\ref{fig:tail_stream}, there is a distinctive peak at the mean RV of the cluster. To select the tidal tails potentially associated with the cluster, we applied selection masks in proper motion, RVs and the CMD simultaneously. Aiming not to bias the selection of extra-tidal stars to have the same kinematics as the cluster stars, we applied relatively wide cuts in each parameter space. 

\begin{figure*}
\includegraphics[width=\textwidth]{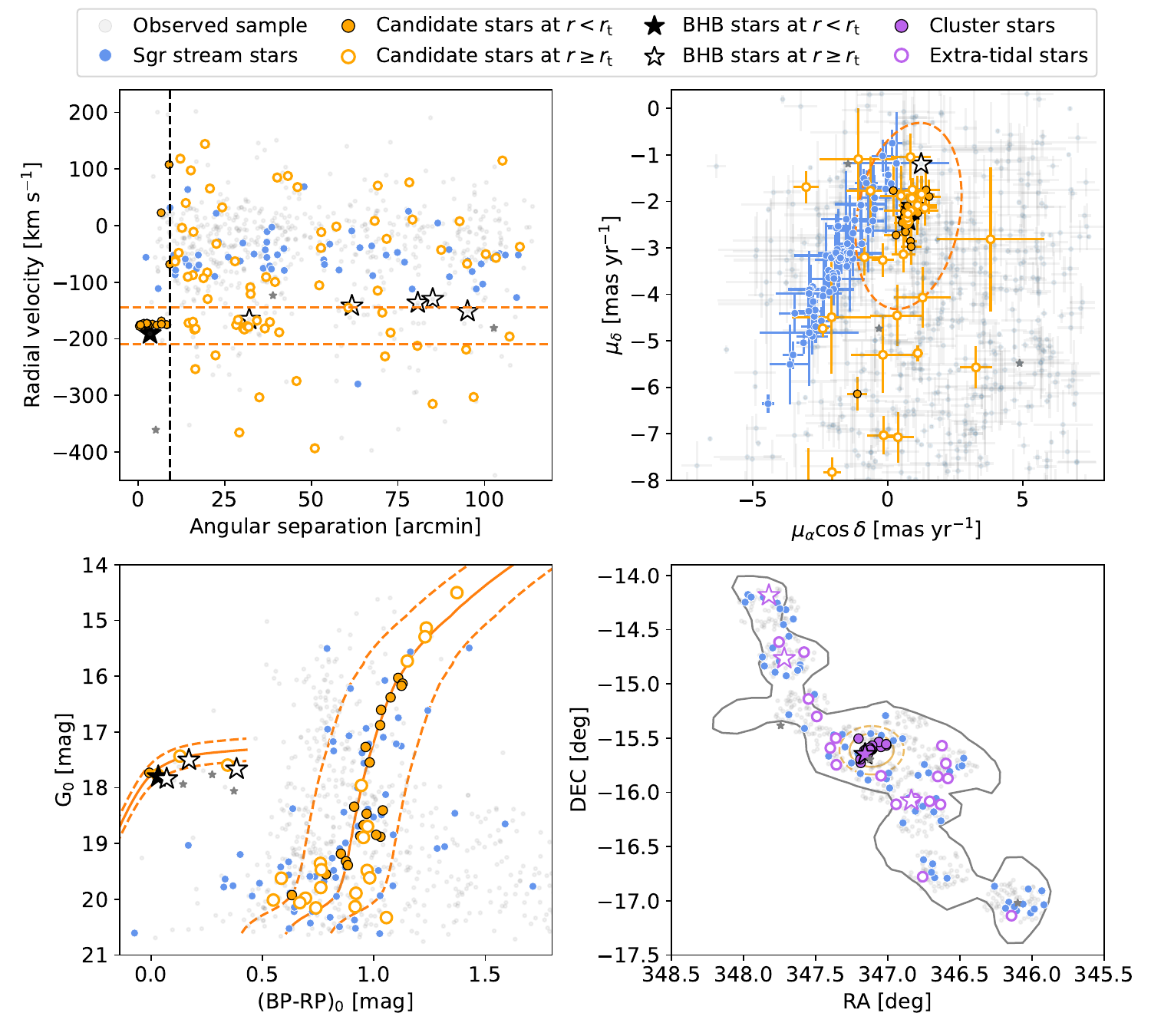}
\caption{Distribution of the observed (grey circles), Sgr stream candidates (blue circles) and extra-tidal star candidates. In each subpanel, the extra-tidal star candidates are selected differently. \textit{Top left:} Angular distance versus RV, showing the selected candidates passing the CMD and PM cuts as orange circles. The vertical line corresponds to the tidal radius. The dashed orange lines define the RV cut. Star symbols correspond to BHB candidates. The filled and open circles correspond to stars inside and outside the tidal radius, respectively. \textit{Top right:} Proper motion distribution for all the observed stars and those extra-tidal candidates, selected based on CMD and RV cuts. The dashed orange ellipse defines the PM cut. \textit{Bottom left:} \textit{Gaia} CMD including all the observed stars, an isochrone (solid orange line) and a tolerance of $\pm$0.2 mag from it (orange dashed lines). \textit{Bottom right:} Spatial distribution of the tidal tails recovered in \citetalias{Navarrete17} (grey contour), the observed stars as grey dots, Sgr stream candidates (blue circles) and the extra-tidal star candidates that pass the cuts in the other three panels as filled and open violet circles and stars (for the BHB stars). The inner and outer orange circles correspond to the tidal and Jacobi radius of the cluster.}\label{fig:masks}
\end{figure*}

The distributions of the observed stars in angular separation versus RV (upper left panel), proper motion (upper right panel), CMD (bottom left panel) and spatial space (bottom right panel) are shown in Figure~\ref{fig:masks} as grey dots. In each panel, the blue dots correspond to the Sgr stream candidate stars, selected as detailed in Section~\ref{sec:sgr}. In all the panels, except the spatial distribution, the filled and open orange circles (inside and outside the tidal radius) are a subsample of stars selected after applying the selection cuts (shown as dashed orange lines) from the other two panels. In particular, the cut in RV corresponds to $\pm$35 km s$^{-1}$ from the mean RV of the cluster $-$176.7 km s$^{-1}$ \citep{BaumgardtHilker18}, wide enough to account for the expected RV variations along the stream track over the extension of the tails ($\approx$4 degrees covered by our observations), the selection in proper motion is 10 times the size of the ellipse obtained from the fit to the proper motion of the stars inside the tidal radius (see Fig.~\ref{fig:tail_stream} and Sec.~\ref{sec:sgr}), equivalent to $\pm$2 mas yr$^{-1}$ with respect to the mean proper motion of the cluster, and the cut in the CMD is restricted to $\pm$0.2 mag (i.e., up to $\pm$2 kpc from the mean distance of the cluster) from a PARSEC \citep{Bressan12} isochrone for the \textit{Gaia} photometry \citep[{[}M/H{]} = --1.7, age = 12 Gyr, E(B-V) = 0.0 mag, see][]{Cohen05, Forbes10, FigueraJaimes13}. These cuts ensure to recover stars in the tails having different kinematics than the cluster, as well as possibly different heliocentric distances and offsets in the isochrone due to inaccurate parameters for metallicity, age and extinction. To remain conservative, the observational uncertainties were not considered when applying the different selection cuts. Stars that passed the three cuts simultaneously are shown as purple filled and open circles in the bottom right panel. Extra-tidal stars are recovered over the entire extension of the proposed tidal tails in \citetalias{Navarrete17}, having more stars associated with the cluster population in the inner region. All the extra-tidal stars (beyond the tidal radius of the cluster, r$_t$ = 9.2 arcmin) are found beyond its Jacobi radius, r$_J$= 13.4 arcmin \citep{Zhang22}. Moreover, Sgr stream stars are recovered as well in all fields, demonstrating that both populations are located along the same are of the sky, despite being physically unrelated.

Regarding the candidate BHB stars (nine sources, shown as star symbols), four of them have RV and proper motions that pass the same cuts as the rest of the candidate cluster stars. They are well located along the horizontal branch (see the bottom left panel of Fig.~\ref{fig:masks}). There are other five BHB candidates that have highly deviating proper motions compared to the rest of the sample (unfilled black stars). In Appendix~\ref{app:bhb}, we listed the BHB candidates as well as one RRc star among this subsample.

Remarkably, applying a selection in RV and a wide selection cut in proper motion produces an almost clean CMD for the cluster stars (bottom left panel), closely following the isochrone of the cluster. Both stars inside and outside the tidal radius are recovered, defining the RGB down to the turn-off point. In the same panel, the distribution of Sgr stream stars has a larger dispersion, but having only the CMD information to select extra-tidal stars would inevitably include Sgr stream stars or field stars. 

\subsection{Metallicity of the tidal tails}

The metallicity content of the extra-tidal stars should be similar to the cluster’s mean metallicity, within the uncertainties. Given the distance of the cluster, our highest S/N observations are restricted to the subsample of red giant branch stars, i.e., a fraction of the full sample. Therefore, even though we derived the metallicity for the full sample, we only considered as reliable those from stars with coadded spectra having S/N$\geq$ 10. We also did not consider the metallicities recovered for BHB stars as they tend to lie at the end of the grid (see Appendix~\ref{app:bhb}). The top panel in the Figure~\ref{fig:meh_ca_ferre} shows the RV and metallicity, only for the measurements with errors in [M/H]$_{\rm FERRE} \leq$ 0.1 dex, and S/N $\geq$ 10. From the 683 stars passing the cut in S/N, having reliable RV measurements, 484 of them passed the cut in metallicity error. The vertical dashed line corresponds to the mean RV of the cluster, while the two horizontal lines are the metallicities from \cite{Cohen05} and \cite{FigueraJaimes13}, respectively (see Appendix~\ref{app:lit_values} for a summary of literature values). The stars that passed all the cuts in RV, CMD and proper motion are marked with unfilled and filled purple symbols if they are inside or outside the Jacobi radius, respectively (19 stars out of 20 and eleven stars out of 18, respectively). There is only one extra-tidal candidate\footnote{\textit{Gaia} source ID 2410110457085084288.} that has a [M/H] = --0.73, incompatible with being associated with the cluster population. Therefore, we exclude it as a possible extra-tidal star for the rest of the analysis. Sgr stream candidates, selected based on proper motions, are shown with blue circles. It is clear that the stars associated with the GC form a relatively compact group at the mean RV of the cluster, although with a relatively large metallicity dispersion.

The bottom panel of Fig.~\ref{fig:meh_ca_ferre}, which shows the metallicity as a function of the S/N ratio, demonstrates that the apparent dispersion in metallicity for the cluster stars is most likely due to the relatively low SNR of the spectra. For stars having larger S/N, the metallicity values are almost all consistent with the two reference values from the literature. There are three more stars with relatively low metallicity ([M/H]$\approx$--1.7) having similar RV to the cluster, but that were not selected as associated with it. We inspected these cases individually and found that they have largely different proper motions and are almost 2 mag brighter than the isochrone for the cluster, at their color, which would imply stars that are located at $\approx$ 10 kpc, in the foreground. Therefore, we excluded them as potentially associated to the cluster, even though they are metal-poor stars with similar RV as the cluster.

The metallicities recovered for the stars inside and outside the cluster tidal radius are fully compatible, having a median value of --1.89 and --1.78, respectively, with dispersions of 0.18 and 0.39 dex. The Sgr stream candidate stars have much higher metallicities, with a median of [M/H]=--1.03 and standard deviation of 0.6 dex, consistent with previous measurements in the literature \citep[{[}M/H{]}$\approx$--1.0, see][]{Cunningham24}. Nonetheless, some of the most metal-rich Sgr stream candidate stars could have a metallicity more compatible with being disk stars instead of part of the the Sgr stream.

\begin{figure}
\includegraphics[width=0.5\textwidth]{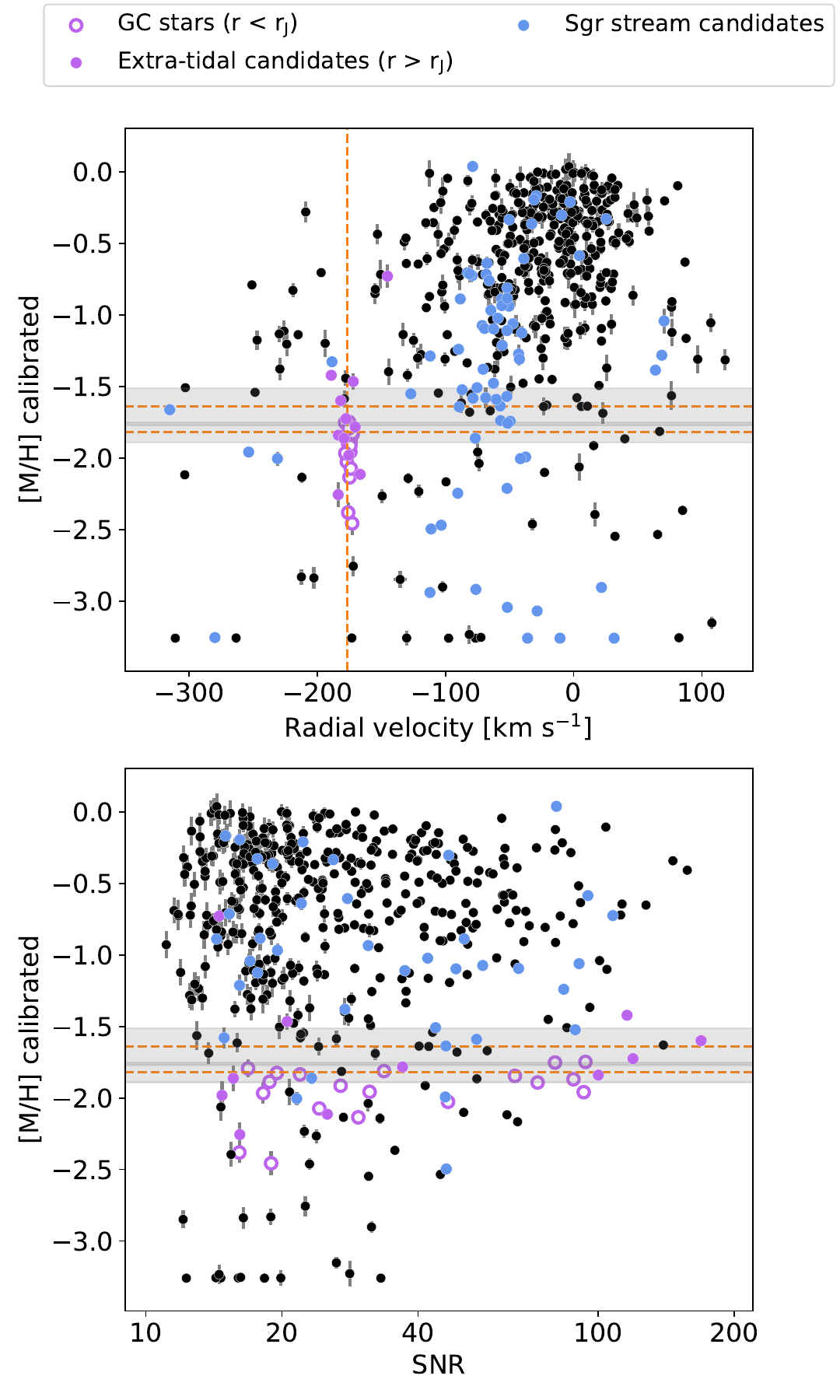}
\caption{Metallicity from the FERRE analysis of the spectra. Top panel: RV and metallicity for a subsample of stars, having spectra with S/N $\geq$ 10 and [M/H] errors$\leq$0.1 dex. Stars associated to the cluster (inside and outside the Jacobi radius) are marked with violet unfilled and filled circles, respectively, while stars likely associated with the Sgr stream are marked with blue circles. The horizontal lines correspond to metallicities of --1.82 and --1.64, with the shaded grey areas corresponding to the metallicity uncertainties, and the vertical line is the mean RV of the cluster. Bottom panel: S/N ratio and metallicity for the same sample.}\label{fig:meh_ca_ferre}
\end{figure}

Table~\ref{tab:extra} in Appendix~\ref{app:extra_tidal} lists the selected cluster stars inside and outside its tidal radius (20 and 18 stars, respectively), including the \textit{Gaia} DR3 \texttt{source\_id}, RVs, [M/H] (for those with S/N$\geq$ 10), S/N ratio and the angular separation (in arcmin) from the center of the cluster. In Appendix~\ref{app:lit_values}, the mean RV and [M/H] for the cluster's stars were derived and compared with previous values in the literature. We estimated a median RV = $-174.9 \pm 0.5$ km s$^{-1}$, with a velocity dispersion of 1.6$\pm$0.3 km s$^{-1}$ and a median [M/H] = $-$1.89$\pm$0.11 dex. Although the analysis of the stars in the cluster is beyond the scope of this paper, the lack of abundant measurements in the literature makes these estimates relevant for future works.

\section{Discussion}\label{discussion}

\subsection{Simulated tidal tails} \label{sim}

To asses the reliability of the detected tidal tails, we derived the expected spatial distribution of extra-tidal stars based on the particle spray models of \cite{Chen25}, as implemented in \texttt{gala} v1.9.1 \citep{Price-Whelan2017}. The \texttt{MilkyWayPotential} consisting of dark matter halo following a Navarro-Frenk-White potential, with a halo mass of M$_h$ = 5.4 $\times$ 10$^{11}$ M$_{\odot}$ and a scale radius r$_s$ =~15.62 kpc, a Miyamoto-Nagai potential for a disk \citep[based on the model of][]{Bovy15} of mass M$_d$ = 7.5 $\times$ 10$^{10}$ M$_{\odot}$, a scale length $a$~=~3 kpc, and a scale height $b$ = 280 pc, plus a spherical Hernquist nucleus and bulge components, with a mass and concentration of M$_n$ = 1.71 $\times$ 10$^{9}$ M$_{\odot}$, $c$ = 0.07, and M$_b$~=~5.0 $\times$ 10$^{9}$ M$_{\odot}$, $c$ = 1.0 kpc, respectively, were adopted. The potential of the cluster was also considered in the model. A Plummer potential having a present-day mass of M$_{\rm GC}$ = 2.0 $\times$ 10$^{4}$ M$_{\odot}$ \citep{VasilievBaumgardt21} and a scale radius of $b \approx r_{h,m}/1.305 =$ 8.11 pc for a half-mass radius of $r_{h,m}$ = 10.59 pc \citep{VasilievBaumgardt21}.

\begin{figure}
\includegraphics[width=0.5\textwidth]{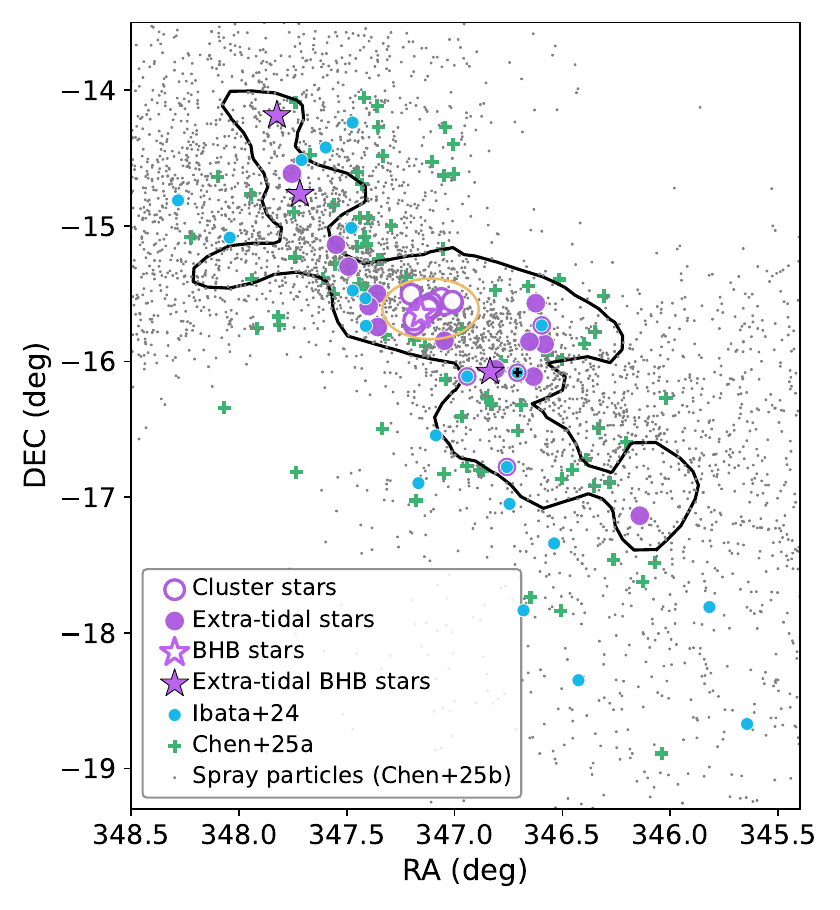}
\caption{Spray particle distribution for the cluster stars \citep{Chen25}. The NGC~7492 cluster and extra-tidal stars are shown as purple circles (or purple stars for BHB stars), while the stars from the stream \#84 from \citet{Ibata24} and the stars from \cite{Chenstreams} are shown as light blue circles and green crosses, respectively. The inner orange circle corresponds to the tidal radius of the cluster. The contour of the original tidal tails detected in \citetalias{Navarrete17} is shown in black. } \label{fig:df}
\end{figure}

Figure~\ref{fig:df} shows the spatial distribution of spray particles obtained using the distribution function from \citet{Chen25}. The distribution of the cluster and extra-tidal stars identified in this work is shown in purple, while the extra-tidal star candidates from \cite{Ibata24} and \cite{Chenstreams} are shown as light blue circles and green crosses, respectively. Among them, only one star has a previous RV measurement in the literature (marked with an inner black plus symbol\footnote{RV = --178.5 $\pm$ 9.7 km s$^{-1}$ from \textit{Gaia} DR3, source ID 2409579805286184448.}), and three of them seem to overlap with three of our targets (light blue symbols with purple borders); however, the angular separation is significantly larger than 1 arcsec. Therefore, there are no stars in common between our sample and the one from \cite{Ibata24}. We note that, as already mentioned in \citetalias{Navarrete17}, there are more BHB stars towards the northern tail of the cluster compared to the southern tail. There are 18 stars in common between our sample and those reported in \cite{Chenstreams}, although only eight are inside our extra-tidal candidate stars. None of them had previous RV measurements. In Appendix~\ref{app:chen_stream}, the cross-match between the stars from \cite{Chenstreams} and our sample is shown, demonstrating that the RVs and metallicities of some of the high-probability candidates are not consistent at all with the cluster population. 

The simulated particles reproduce the same overall orientation and extent as the observed extra-tidal population, extending over more than 2 deg from the cluster center. The agreement between the simulations, the previously reported photometric detections from \citetalias{Navarrete17}, and the spectroscopically confirmed extra-tidal stars strongly supports the interpretation that NGC 7492 is surrounded by extended tidal tails. 

A considerable spatial dispersion is recovered both in the simulations and in the candidate tidal-tail stars from different catalogues. Such broad tidal structures are expected for dynamically evolved outer-halo GCs and are commonly associated with accreted systems \citep{Malhan21, Callingham22}. In this context, the morphology of the tidal tails of NGC 7492 is consistent with the cluster being accreted, probably from a different merger event than the Sgr dwarf galaxy, most likely to the Helmi stream \citep[see e.g.,][and references therein]{Koposov12, Callingham22}.

\begin{figure}
\includegraphics[width=0.5\textwidth]{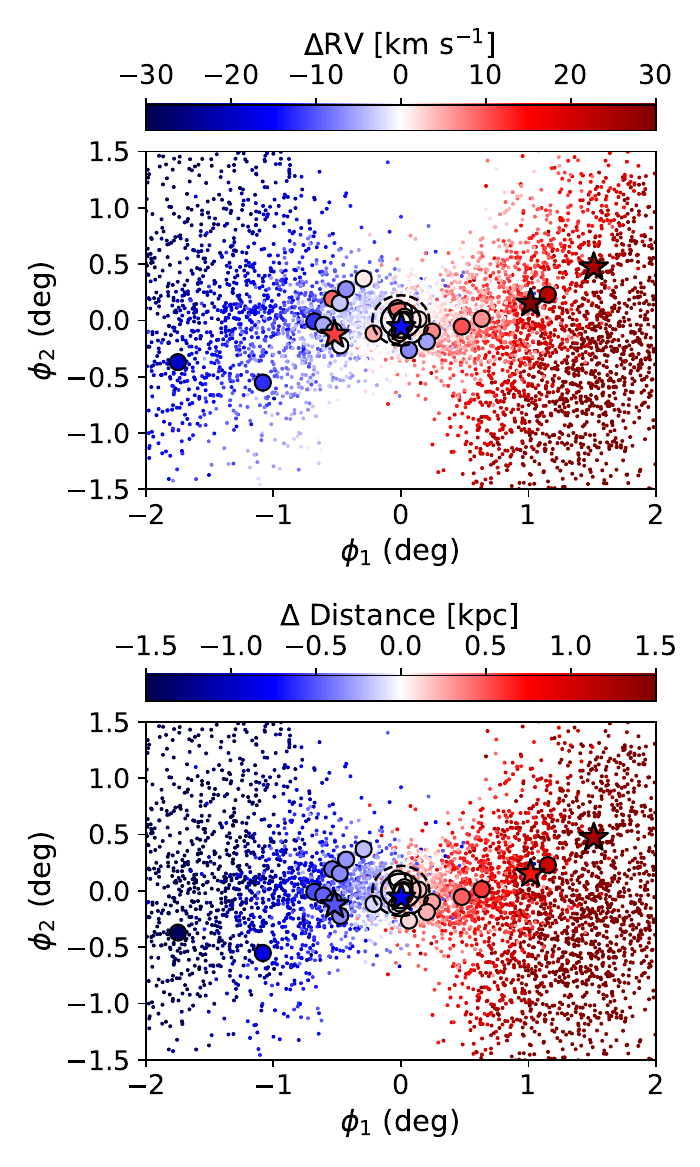}
\caption{RV and distance variations along the tidal tails. Top panel: RV difference with respect to the mean RV of the cluster for the stream particles from the \citet{Chen25} distribution function (background points) and the observed extra-tidal candidates (circles and stars). Bottom panel: Distance difference from the particles and stars.} \label{fig:df_delta}
\end{figure}

For better visualisation, we defined a new coordinate system ($\phi_1$, $\phi_2$) where $\phi_1$ is aligned along the direction of the simulated particles and $\phi_2$ is the orthogonal direction to it. The equations to transform from equatorial coordinates ($\alpha$, $\delta$) to this new coordinate system are presented in the Appendix~\ref{app:rotation}. To compare the RV of the simulated particles and the observations, we used the predicted RVs from the model of \cite{Chen25}. The top panel of Figure~\ref{fig:df_delta} shows the distribution of the RV differences, as the individual RV minus the median velocity of the cluster ($-$176.7 km s$^{-1}$), showing a good agreement between both cluster stars (inside and outside the tidal radius) and the simulated particles. A gradient in RV is seen, covering $\pm 30$ km s$^{-1}$, similar to the range of RV originally imposed to get the extra-tidal candidates ($\pm$35 km s$^{-1}$), see Appendix~\ref{app:rv_gradient} for a more detailed discussion on the velocity gradient. We noted that among the four horizontal branch stars (three BHBs and one RRc variable, as star symbols), one star has a positive $\Delta$RV at $\phi_1 \leq$ 0 degrees, making it less likely to be associated with the extra-tidal population. The RRc star, at positive $\phi_1$, has the largest $\Delta$ RV. Depending on the phase at which the spectra were obtained, the RV of this star can be shifted from the systemic motion of the star due to its variable nature.

A distance gradient is also expected, given the different radial velocities and the orbit of the cluster (see Sect.~\ref{sec:orbit}). However, \textit{Gaia} DR3 parallaxes are not accurate enough, at the distance of the cluster, to be used to get individual and precise distance estimates. Therefore, we measured the distance gradient of the stream using the simulated particles in the stream-aligned coordinate system ($\phi_1$, $\phi_2$). Because the stream is slightly curved even in this rectified frame, and has a considerable width, we modelled its central locus by fitting a smoothing spline $\phi^{\rm spline}_1$($\phi_2$) to the particle distribution. For each particle, we computed the perpendicular displacement $\Delta \phi_2$, and used its 68th-percentile width to obtain a robust empirical estimate of the stream thickness, $\Delta \phi^{\rm 68}_2$. We then selected only particles satisfying $\Delta \phi_1 < \Delta \phi^{\rm 68}_2$, which isolates the central stream while excluding outliers. To measure the distance gradient along the stream, we used the particles satisfying the condition of being inside the 68th percentile width, being  $\Delta \phi^{\rm 68}_2 =$ 0.38 deg. This yields a gradient of dD/d$\phi_1 \approx$ 1.08 kpc deg$^{-1}$. Finally, we applied this empirical gradient to estimate heliocentric distances for the candidate extra-tidal stars using their measured $\phi_1$ values and the fitted gradient. The bottom panel in Fig.~\ref{fig:df_delta} shows the result of this procedure, showing that the simulations predict a distance difference of $\pm2$ kpc from the mean distance of the cluster (24.39 kpc). This is equivalent to a magnitude difference of $\Delta$G $\approx \pm$ 0.18 mag, of the order of our selection cut in the CMD based on the isochrone.

Consistent distances with the rest of the distribution were obtained for the three BHB stars, even though we did not use the simulated particles in this case. For each horizontal branch star, we used the Zero Age Horizontal Branch for the cluster \citep[ZAHB, retrieved from the PGPUC database, adopting Y = 0.23, Z = 0.00059 and {[}$\alpha$/Fe{]} = 0.3 dex, see][]{Aldo25} and the stars inside the tidal radius of the cluster observed in \textit{Gaia} DR3. As stars evolve from the ZAHB, the observed horizontal branch stars of the cluster were slightly brighter. We adjusted the position of the ZAHB to match the observed horizontal branch and used that reference to derive relative distances for the BHB candidates. For the RRc star, we used the Eq.~20 from \cite{Prudil24} and its period (converted to the fundamental mode period) and intensity-average RP magnitude from the \textit{Gaia} DR3 characterization of variable stars \cite{Clementini2023}, while adopting the metallicity of the cluster to get the distance. The two stars that have deviating $\Delta$ RV appear with consistent distance as the rest of the extra-tidal candidates and the simulated particles at the corresponding $\phi_1$ values. The distance of the RRc star, giving its color and magnitude is quite similar to the rest of the stars at positive $\phi_1$. We noted that this cluster contains only one RRc star (and one ab-type RR Lyrae star) as confirmed members inside the tidal radius \citep{Clement2001}. If confirmed, this new star would be the first extra-tidal variable star of this cluster.

\subsection{NGC 7492's orbit}\label{sec:orbit}

\begin{figure}
\includegraphics[width=0.5\textwidth]{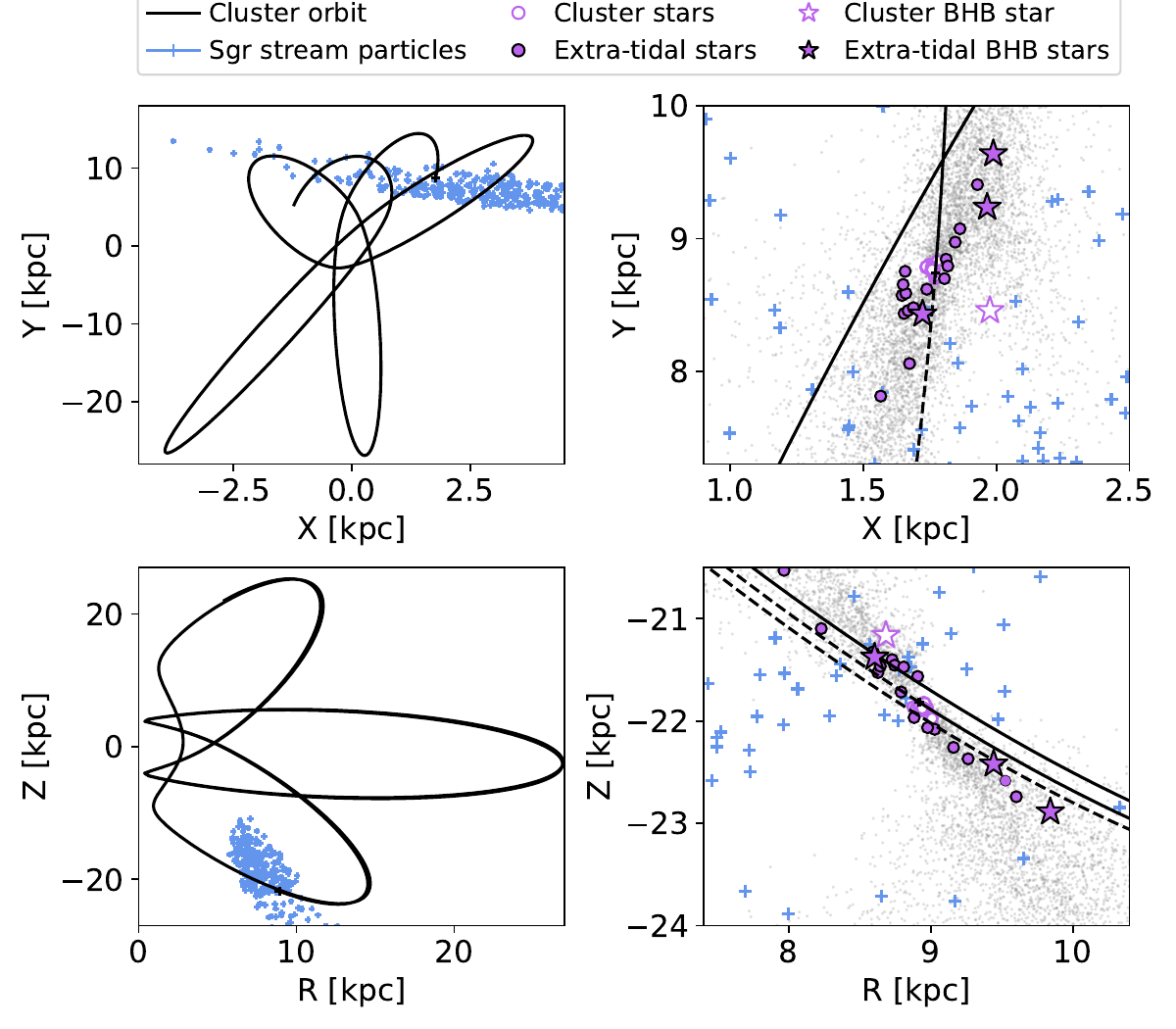}
\caption{Galactocentric coordinates for the integrated orbit of the cluster (black lines). The corresponding distribution of spray particles is included as grey dots, while the cluster and extra-tidal candidate stars adopting the corresponding distance based on the distance gradient are shown violet circles. The BHB stars are shown as star symbols.} \label{fig:orbit}
\end{figure}

Thanks to the \textit{Gaia} DR2 and DR3 astrometric observations \citep{Gaia2023}, the mean proper motion of the cluster was derived \citep{Vasiliev19, BaumgardtHilker18}. To compare the distribution of extra-tidal stars and to analyze the orbit of the cluster, we integrated the present-day 6D phase-space position of NGC 7492 using \texttt{galpy} and the \texttt{MWPotential2014} \citep{Bovy15}, for 1 Gyr back in time. The observed position were transformed to Galactocentric coordinates, using \texttt{astropy} and adopting R$_{\odot}$ = 8.122 kpc, and Z$_{\odot}$ = 20.8 pc \citep{Gravity19, Bennett19}. Figure~\ref{fig:orbit} shows the orbit of the cluster in Galactocentric $X$, $Y$ (top panels) and $R$, $Z$ (bottom panels) as a solid black line. The distribution of the spray model particles from the distribution function of \cite{Chen25} are shown as grey dots in the left panels. The orbit is consistent with a halo GC, having an eccentric ($e\approx$0.8) and almost polar orbit (inclination $\approx$ 97$^{\circ}$), as previously reported in the literature \citep{Armstrong21, Bajkova21} based on \textit{Gaia} DR2 proper motions. The cluster has recently passed apocentre, heading to pericenter. In the right panels, a zoom-in of the orbital path is shown, including the positions of the cluster and extra-tidal stars, adopting the distance gradient found using the spray particle model (violet points). The Sgr stream particles from the simulation of \cite{Vasiliev21} up to 5 degrees from the cluster's center are shown as blue crosses in all the panels. The inner extra-tidal stars and spray particles seem to follow the orbit, as expected for a cluster heading to pericenter \citep[see e.g., NGC 1261 in][]{Awad25}, while, at larger distances there seems to be a more significant offset in the trailing tail compared to the orbit, which is not seen in the leading tail, both in the simulated particles as in our extra-tidal stars. 

\subsection{Implications for the origin of NGC~7492}

Although NGC 7492 spatially overlaps with the Sgr stream, the distinct RVs and proper motions of the cluster and its extra-tidal stars clearly demonstrate that the system is not associated with the debris from Sgr. The overlap between both structures is only a projection effect, with the cluster and the Sgr stream sharing similar heliocentric distances while exhibiting markedly different kinematics. The similar positions and distances tend to mix the Sgr and cluster population in photometric studies, being only unequivocally separated using kinematics.

While the present-day kinematics clearly rule out an association between NGC 7492 and the trailing arm of the Sgr stream, a more ancient connection with the Sgr disruption cannot be entirely excluded. In particular, if the cluster was accreted during an early passage of Sgr, several Gyr ago, the phase mixing could erase a direct correspondence with the present-day stream debris. However, the current orbital and kinematic properties of NGC 7492 do not provide compelling evidence for such a scenario, as Sgr models \citep{Vasiliev21} predict a much different RV signature for the older Sgr arm, at the position of the cluster (RV$\sim-$270 km s$^{-1}$, see Fig.~\ref{fig:lambda_vgsr}).

The lack of association with the Sgr stream does not argue against an accreted origin for NGC~7492. On the contrary, the orbital properties of the cluster, including its eccentric and nearly polar orbit, together with the broad morphology of its tidal tails, are consistent with expectations for accreted outer-halo GCs. Previous works have indeed suggested that NGC~7492 is more likely associated with the Helmi streams rather than with Sgr \citep{Callingham22}. The confirmation of extended tidal tails around NGC 7492, therefore, provides additional evidence that dynamically disrupting accreted GCs could contribute stars to the assembly of the outer Galactic halo.

Based on the kinematics of the tidal tails and the cluster's orbit, the most plausible scenario for the origin of NGC~7492 and its tails is that the cluster was accreted at early times, between 6 and 9 Gyr ago if associated with the Helmi stream, and is currently undergoing tidal disruption in the Galactic halo. The ongoing tidal stripping of the cluster is naturally expected from its eccentric orbit and repeated pericentric passages within the MW potential, while the comparatively broad morphology of the tidal tails may reflect the dynamical history of the system prior to or during its accretion onto the MW, as suggested for other accreted outer halo GCs \citep[see e.g.,][]{Kuzma2018,Malhan2019, Qian22}. If NGC 7492 was accreted together with the Helmi progenitor, this survival time is not unusual for Galactic GCs, many of which have remained bound for a Hubble time despite continuous tidal mass loss. Moreover, the inferred accretion time does not necessarily correspond to the time spent by NGC~7492 on its present-day orbit, since the cluster may have remained bound to its progenitor for part of its evolution before being released into the Galactic halo as the dwarf galaxy was disrupted.

\section{Summary and conclusions}\label{summary}

In this work, an extensive spectroscopic follow-up of 4.0 degrees along the expected tidal tails of NGC 7492 was performed. FLAMES high-resolution multi-object spectra were obtained for 766 of stars in ten fields, aiming to identify the extra-tidal stars of the cluster as well as Sgr stream stars, expected to be at the same distance as the cluster. From the spectra, reliable RVs were derived for 718 stars of which 484 have reliable metallicity measurements (S/N $\geq$ 10, [M/H] uncertainty $\leq$ 0.1 dex).

To isolate Sgr stream stars, a proper motion selection was performed, based on the predicted proper motions for the particles in the model of \cite{Vasiliev21}. The RVs of these stars present one prominent peak, at $\approx$--80 km s$^{-1}$ which corresponds to the trailing arm of the Sgr stream, while a secondary, less prominent peak at $\approx$80 km s$^{-1}$ is likely associated with the leading arm. These stars are located between 2.3 arcmin (i.e., inside the cluster's tidal radius) up to 1.8 deg from the cluster center, with a distinctive mean metallicity of [M/H] = --1.03. 

To identify stars associated with the cluster, we performed a simultaneous cut in RV, proper motions and the difference in color and magnitude with respect to an isochrone for the cluster. The cuts in RV and proper motion were wide enough to recover stars with slightly different kinematics than the cluster. Our selection recovered 20 stars inside the Jacobi and tidal radii of the cluster and 17 stars outside the Jacobi radius, as far as 1.78 deg from the cluster center. Although the tidal tails of the cluster have a relatively low-density of stars, the kinematic signature of the cluster is recovered, confirming the early photometric detection of tidal tails by \citetalias{Navarrete17}. The metallicities of the extra-tidal stars are consistent with the metallicities of the stars inside the tidal radius, and clustered around the mean value of the cluster [M/H]$\sim$--1.89 dex.

We used numerical simulations and the orbit of the cluster to interpret the spatial and kinematic distribution of the extra-tidal stars, finding that the variation in RV and distance along the tails can be as large as $\pm$30 km s$^{-1}$ and $\pm$2 kpc, respectively. In particular, we derive a distance gradient of 1.08 kpc deg$^{-1}$. We recovered two BHB stars and one RRc variable star among the extra-tidal population. The extra-tidal stars seem to follow the same spatial distribution as the simulation particles while slightly deviating from the cluster's orbit. 

This work provides the first extensive spectroscopic confirmation of the tidal tails of NGC 7492 and confirms the coexistence of Sgr stream stars and extra-tidal cluster stars at the same distance and sky position, but with clearly distinct kinematics and metallicities. In particular, this work highlights how tidal tails associated with GCs can remain identifiable even in regions strongly contaminated by major halo substructures such as the Sgr stream. The present dataset offers significant potential for future exploitation, including improved metallicity determinations for faint stars through spectral resampling and CaT analyses, enabling a more complete characterization of the stellar populations associated with the tidal tails. Future spectroscopic observations, including high-resolution, high S/N spectra, could extend the present analysis to the numerous faint main-sequence population, significantly increasing the number of stars tracing the tidal tails of this cluster.

\begin{acknowledgements} We warmly thank the referee for an insightful referee report. C.N acknowledges financial support from the Centre National d'Études Spatiales (CNES) fellowship program. C.N. acknowledges Sof\'ia Gran Navarrete for her contribution to the design of the plots. A. R. A. acknowledges support from DICYT through grant 062319RA. S.K. acknowledges support from the Science \& Technology Facilities Council (STFC) grant ST/Y001001/1. V.B. is grateful for the support from the Leverhulme Trust Research Project Grant RPG-2021-205 ‘The Faint Universe Made Visible with Machine Learning’. P. B. acknowledges funding from the CNES post-doctoral fellowship program. E. V. acknowledges funding from the Royal Society, under the Newton International Fellowship programme (NIF\textbackslash R1\textbackslash 241973). This work has made use of data from the European Space Agency (ESA) mission \textit{Gaia} (https://www.cosmos.esa.int/gaia), processed by the \textit{Gaia} Data Processing and Analysis Consortium (DPAC, https://www.cosmos. esa.int/web/gaia/dpac/consortium). Funding for the DPAC has been provided by national institutions, in particular the institutions participating in the \textit{Gaia} Multilateral Agreement. This research made use of \texttt{Astropy}, a community-developed core Python package for Astronomy \citep{astropy13, astropy18, astropy22}, of the \texttt{galpy} package for Galactic dynamics \citep{Bovy15}, and of the \texttt{gala} package for Galactic dynamics and orbit integration \citep{Gala17}. 
\end{acknowledgements}

\bibliographystyle{aa}
\bibliography{biblio}

\begin{appendix}

\onecolumn{
\section{GIRAFFE data}
\subsection{RV comparison with the literature}\label{app:rv_comp}

Given the faint magnitude of the observed stars, there are very few literature RV measurements to compare with. We have found six stars in our sample with previous RV measurements available in \textit{Gaia} DR3 and four stars with measurements from Keck/DEIMOS spectra reported in \cite{Geha26a} (see Fig. \ref{fig:rv_gaia_flames}). The two datasets have a good overall agreement with the measurements from this work. The median RV difference for the ten stars is RV$_{\rm Lit}$ - RV$_{\rm GIRAFFE}$ = 0.85 km s$^{-1}$, with a robust MAD of 2.40 km s$^{-1}$. This indicates no significant systematic offset in the RV estimation, given the small sample size and the scatter of the comparison.

We also estimated the median normalized difference as

\begin{equation}
    \frac{({\rm RV}_{\rm Lit} - {\rm RV}_{\rm GIRAFFE})}{\sqrt{\sigma_{\rm RV_{\rm Lit}}^2 + \sigma_{\rm RV_{\rm GIRAFFE}}^2}} \text{,}
\end{equation}

obtaining a value of 0.4, corresponding to a small offset of 0.4$\sigma$ between the two RV scales.

The robust MAD of the normalized difference is 0.7, suggesting that the adopted RV errors may be overestimated, being a conservative estimate. In our analysis, the typical RV uncertainty is of the order of $\sim$0.5 km s$^{-1}$, which is relatively large compared to the expected precision of the GIRAFFE HR21 setup.

\begin{figure}[h!]
    \centering
    \includegraphics[width=0.5\textwidth]{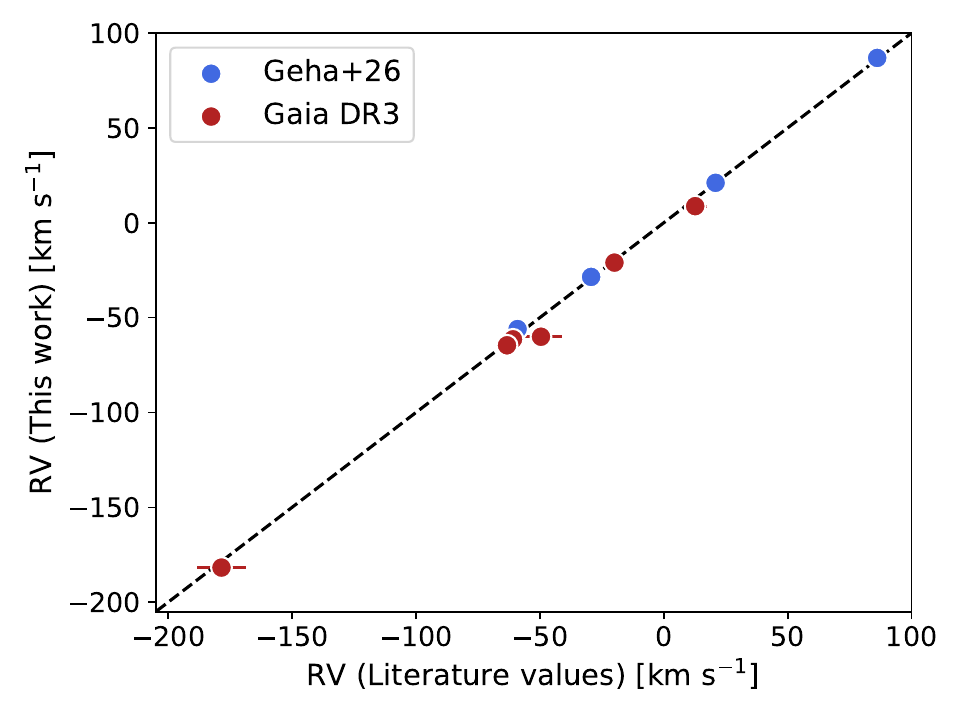}
    \caption{RV comparison with the values from the literature. Six stars in common between our full sample of more than $\sim$700 stars and the \textit{Gaia} DR3 and four stars in common with the sample of \cite{Geha26a}. The black dashed line corresponds to the 1:1 relation.}
    \label{fig:rv_gaia_flames}
\end{figure}

\subsection{SNR threshold}\label{app:snr}

To assess the quality of the RV and metallicity determination, we have compared their uncertainties with the SNR and magnitude of the observed stars. The derived RV and metallicity errors as a function of the $G$ magnitude and S/N ratio are shown in Figure~\ref{fig:sn}. Most of the targets are fainter than G = 19 mag, the full sample having a median of $G$=19.2 mag and S/N$\sim$12, RV error of 0.5 km s$^{-1}$ and median [M/H] error of 0.06 dex. The [M/H] errors tend to significantly increase for stars having S/N $<$ 10, the adopted threshold below which the data is not considered for the analysis. This cut discards 304 stars of the original 766 stars observed. For RV errors, only sources having an error in RV $<$ 10 km s$^{-1}$ were considered (i.e., discarding 48 stars with highly inaccurate measurements).

\begin{figure}[h!]
\includegraphics[width=\textwidth]{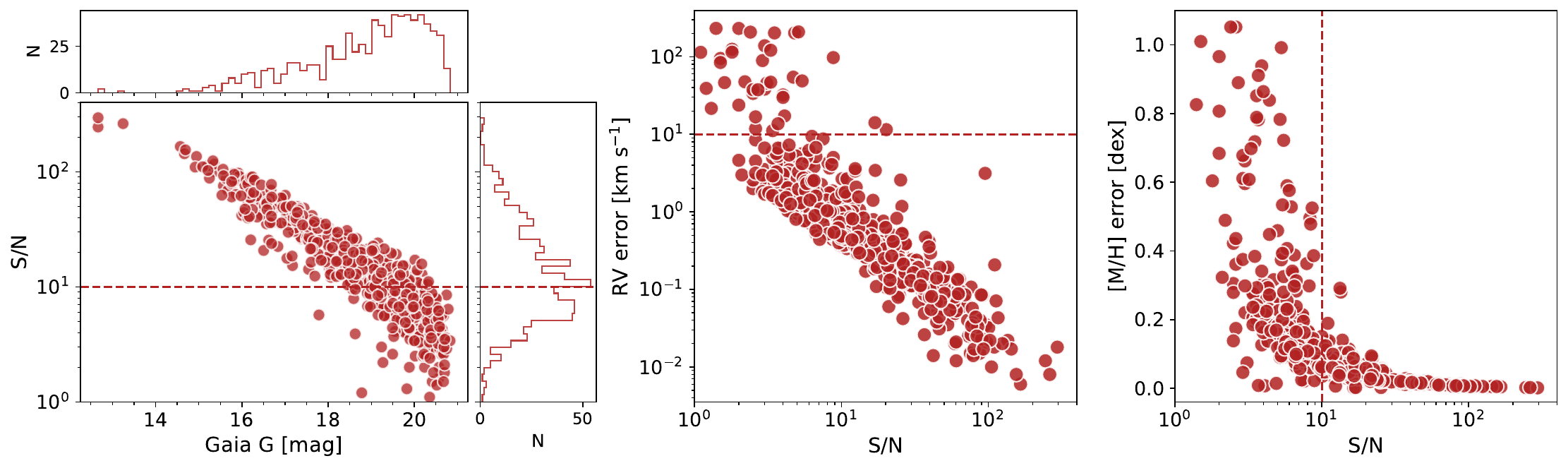}
\caption{Mean S/N from the co-added spectra versus the $G$ magnitude of the source (left panel). Middle and right panels show the RV error and [M/H] uncertainty as a function of the magnitude $G$.}\label{fig:sn}
\end{figure}

\newpage
\section{Sgr stream candidate stars}\label{app:tbc}

In Table~\ref{tab:sgr_stream}, the Sgr stream candidates, selected based on their proper motions (see top panel in Fig~\ref{fig:tail_stream}), are presented. The \textit{Gaia} DR3 source ID, RV, metallicity (derived from spectra having S/N$\geq$10, angular distance from the cluster are included. Those stars listed as Sgr stream candidates in \cite{Ramos22}, all of them without RV values in the literature, are marked in Remarks.  

\begin{longtable}{cccccccc}
\caption{Sgr star candidates, selected based on their \textit{Gaia} proper motions.\label{tab:sgr_stream}}\\
\hline
 \textit{Gaia} DR3 source\_id  & RV  & $\sigma_{\rm RV}$ & [M/H] & $\sigma_{\rm [M/H]}$ & SN & r & Remarks \\
                    & (km s$^{-1}$) & (km s$^{-1}$) &     & & & (arcmin)    & \\
\hline
\endfirsthead

\caption[]{Sgr star candidates (continued)}\\
\hline
 \textit{Gaia} DR3 source\_id  & RV  & $\sigma_{\rm RV}$ & [M/H] & $\sigma_{\rm [M/H]}$ & SN & r & Remarks\\
                    & (km s$^{-1}$) & (km s$^{-1}$) &     & & & (arcmin)   & \\
\hline
\endhead

\hline
\endfoot

\hline
\endlastfoot 
2409630893921997056 &  --56.0 & 0.1 & --0.93 & 0.02 &  25 & 2.3 & In \cite{Ramos22}\\ 
2409617768501872768 & --111.2 & 0.1 & --2.50 & 0.02 &  42 & 5.8  & In \cite{Ramos22}\\ 
2409635669925700352 &  --68.2 & 0.7 &        &      &   9 & 9.1 & \\
2409635051450380288 &  --52.2 & 5.1 &        &      &   6 & 10.3 & \\
2409613233016338944 &  --87.0 & 0.1 & --1.52 & 0.01 &  85 & 10.7 & In \cite{Ramos22}\\ 
2409993771413961088 &  --78.8 & 0.1 &  0.04 & 0.01 &  78 & 10.8 & In \cite{Ramos22}\\ 
2409610759115103616 &  --59.1 & 0.2 & --1.02 & 0.01 &  38 & 13.6 & \\
2409634639133532288 &  --66.0 & 0.9 &        &      &   8 & 14.0 & \\
2409608731890543488 &    63.8 & 1.1 &        &      &   8 & 14.3 & \\
2409996893855064832 &  --70.6 & 0.3 & --1.38 & 0.03 &  21 & 14.7 & In \cite{Ramos22}\\ 
2409604505642684288 & --253.5 & 1.8 &        &      &   6 & 16.5 & \\
2409605295916673920 &  --76.3 & 3.3 &       &     &   4 & 16.9 & \\ 
2409996275379771648 &  --51.8 & 0.7 & --0.88 & 0.08 &  11 & 17.4 & In \cite{Ramos22}\\ 
2406986912055065856 &  --69.6 & 0.1 & --1.09 & 0.01 &  62 & 17.5 & In \cite{Ramos22}\\ 
2410060906047722112 &  --82.8 & 2.2 &        &       &   4 & 19.9 & \\  
2409602242195350016 &  --89.8 & 0.1 & --1.24 & 0.01  &  80 & 22.1 & \\
2409623953254696960 &    21.7 & 8.7 &         &      &   8 & 22.6 & \\ 
2409619589567856128 & --28.5 & 2.3 &         &         & 7 & 23.2 & \\
2409597156954093568 &  --55.7 & 0.8 &         &      &   9 & 26.0 & In \cite{Ramos22}\\ 
2410001051383428480 &  --62.6 & 0.8 &        &       &   7 & 29.5 & In \cite{Ramos22}\\ 
2409578632759604992 &  --30.8 & 0.4 & --0.19 & 0.08 &  10 & 30.3 & \\
2409599420401455872 &  --50.3 & 0.2 & --0.33 & 0.04 &  20 & 33.5 & \\
2409646768120959616 &  --75.4 & 0.1 & --1.51 & 0.02 &  39 & 36.1 & In \cite{Ramos22}\\ 
2409576433736323328 &  --42.9 & 2.4 &        &       &   7 & 36.7 & \\ 
2409575024987074944 &  --89.3 & 3.0 &        &       &   4 & 37.1 & \\ 
2409652407413404416 &  --29.3 & 0.4 & --0.17 & 0.05 &  10 & 37.2  & \\
2409570626940538880 & --56.1 & 4.3 & --0.89 & 0.09 & 8 & 37.2 & \\
2409569463004932480 & --71.4 & 0.1 & --1.07 & 0.01 & 51 & 38.0 & In \cite{Ramos22}\\ 
2410046646756434816 & --76.7 & 0.4 & --1.86 & 0.04 & 19 & 38.5 & In \cite{Ramos22} \\
2410025893474013184 & --2.8 & 0.2 & --0.21 & 0.03 & 18 & 38.6 & \\
2406562603645443584 & --40.5 & 0.3 & --1.12 & 0.05 & 13 & 42.1 & \\
2409570150199689472 & --79.2 & 0.1 & --0.72 & 0.01 & 105 & 42.2 & In \cite{Ramos22}\\
2409567809441896320 &    68.8 & 2.2 &       &      & 5 & 47.9 & \\
2410058294707175680 & --41.5 & 1.2 &        &      & 7 & 53.3 & \\
2410056958972899840 & --60.5 & 0.1 & --1.59 & 0.02 & 50 & 53.6 &  In \cite{Ramos22}\\
2410105298829277568 & --68.6 & 1.5 &        &      &  9 & 53.8 & In \cite{Ramos22}\\
2410105711146155008 & --32.7 & 0.5 & --0.36 & 0.06 & 14 & 55.3 & \\
2410054583855420672 & --78.5 & 4.2 &        &      &  5  & 58.0 & \\
2410152788283142016 & --9.4 & 0.1 & --0.30 & 0.02 & 42 & 61.7 & In \cite{Ramos22}\\
2406450109861734912 & --39.0 & 0.3 & --0.6 & 0.03 & 22 & 64.0 & \\
2410153784715552000 & --88.4 & 0.1 & --0.89 & 0.01 & 46 & 67.7  & In \cite{Ramos22}\\
2410202266306394240 & 70.5 & 3.4 & --1.04 & 0.09 & 12 & 69.2 & \\
2410207729504801664 & --231.0 & 0.7 & --2.0 & 0.05 & 16 & 71.3 & \\
2406442310201094784 & --51.9 & 0.2 & --1.11 & 0.02 & 33 & 71.6 & In \cite{Ramos22}\\
2406438912882539904 & --47.4 & 0.1 & --1.06 & 0.01 & 87 & 74.9 & In \cite{Ramos22}\\
2406440489135535360 & --64.6 & 0.4 & --0.97 & 0.05 & 13 & 76.9 & In \cite{Ramos22}\\
2410226382548016128 & 25.4 & 0.4 & --0.33 & 0.04 & 12 & 78.0 & \\
2410227550779136256 & --112.1 & 7.3 &      &     & 4  & 79.1 & \\
2410229891536073856 & --315.1 & 2.6 &      &     & 4 & 85.0 & \\
2410229990320278784 & 4.5 & 0.1 & --0.58 & 0.01 & 93 & 86.5 & \\
2410233533668403840 & --42.3 & 2.5 &        &     & 6   & 89.7 & \\
2406427879111518208 & --90.4 & 0.7 &        &     & 15 & 89.9 & \\
2410254591892974208 & --62.4 & 0.1 & --1.1 & 0.01 & 44 & 95.1 & In \cite{Ramos22}\\
2410243047021442304 & --81.1 & 0.5 & --0.71 & 0.06 & 10 & 96.4 & In \cite{Ramos22}\\
2410255141648788352 & --112.0 & 6.7 &       &      & 3   & 97.6 & \\
2406424099540651264 & --67.7 & 0.3 & --0.64 & 0.05 & 15 & 99.0 & In \cite{Ramos22}\\
2406418601982269312 & --50.5 & 0.9 &        &      & 5 & 100.4 & \\
2406406369915164032 & --57.3 & 0.1 & --1.64 & 0.01 & 41 & 102.4 & In \cite{Ramos22}\\
2409426212960493312 & --57.1 & 4.1 &        &      & 5 & 103.3 & \\
2406418945579435776 & --49.5 & 1.1 &        &      & 8 & 104.4 & \\
2406417605552509952 & --51.9 & 0.7 &        &      & 7  & 106.0 & \\
2400413997184335744 & --127.1 & 1.2 &       &      & 7 & 109.2 & In \cite{Ramos22}\\
2400417677971200512 & --51.8 & 3.5 &        &      & 3 & 109.7 & \\
2400413279924461312 & --37.4 & 0.1 & --1.99 & 0.02 & 40 & 110.2 &  \\

\end{longtable}
}

There are four stars from the sample of \cite{Ramos22} that have RV measurements derived from our spectra but that are not included in the proper motion selection to define Sgr stream's stars. Their slightly deviating proper motions, with respect to the rest, produced that they were not included in the selection. They are listed in Table~\ref{tab:extra_sgr}. In the case of \textit{Gaia} DR3 source ID 2409573650597520512, it was found to be associated with the cluster and therefore it is included in Table~\ref{tab:extra}. 

\begin{table*}[h!]
    \centering
    \caption{RV measurements for three stars listed in \cite{Ramos22} as Sgr stream candidates that were not selected based on the proper motion cut applied to our sample.}
    \begin{tabular}{ccccccc}
   \hline
 \textit{Gaia} DR3 source\_id  & RV  & $\sigma_{\rm RV}$ & [M/H] & $\sigma_{\rm [M/H]}$ & SN & r \\
                    & (km s$^{-1}$) & (km s$^{-1}$) &     & & & (arcmin)   \\
\hline
2406595623354353664 & --49.7 & 0.6 &      &      & 13 & 25.6 \\
2410046509317466240 & --154.7 & 0.3 & --0.82 & 0.03 & 22 & 36.9 \\
2410110968186173568 & --212.4 & 0.8 &     &      & 10 & 58.7 \\
\hline
    \end{tabular}
    \label{tab:extra_sgr}
\end{table*}

\newpage

\section{BHB stars and RR Lyrae stars} 
\label{app:bhb}

From the position in the CMD, nine stars were identified as horizontal branch candidates (referred as BHB candidates through the text due to the significantly blue horizontal branch of the cluster). In particular, these stars have \textit{Gaia} colors in the range of --0.1~$\leq$~(BP-RP)$_0~\leq$ 0.4 mag and 17.3 $\leq$ G$_0 \leq$ 18.0 mag. All but one of them are also included as potential BHB candidates in the sample of \cite{Culpan2021}. 

Among them, we recovered the c-type RR Lyrae (RRc) variable star V3 \citep{Clement2001}\footnote{Here we used the online version of the catalog, last updated in 2024: \url{https://www.astro.utoronto.ca/~cclement/cat/C2305m159}.}, for which there is no data regarding its membership to the cluster. Based on its proper motion and RV, we can confidently discard it as a member of the cluster, even though it is located inside its tidal radius. To estimate its heliocentric distance, the period from \cite{Clement2001} and the G$_{\rm RP}$ magnitude were considered, as this star is not included in the sample of \textit{Gaia} DR3 RR Lyrae. The distance recovered confirms that V3 is a foreground RRc star.

Based on the \texttt{VARIABITLITY\_FLAG} in the \textit{Gaia} DR3, we also found another RRc candidate (\textit{Gaia} source\_id 2410199994268808064) having a period of $P$ = 0.27958 d. It is located more than 1 degree away from the cluster, and therefore, it has not been associated with it previously in the literature. Its proper motion and RV are consistent with the expectations for stars located at such large angular distances (i.e., the RV difference with respect to the cluster mean RV is of the order of 30 km s$^{-1}$). 

\begin{figure}[!h]
\centering
\includegraphics[width=0.45\textwidth]{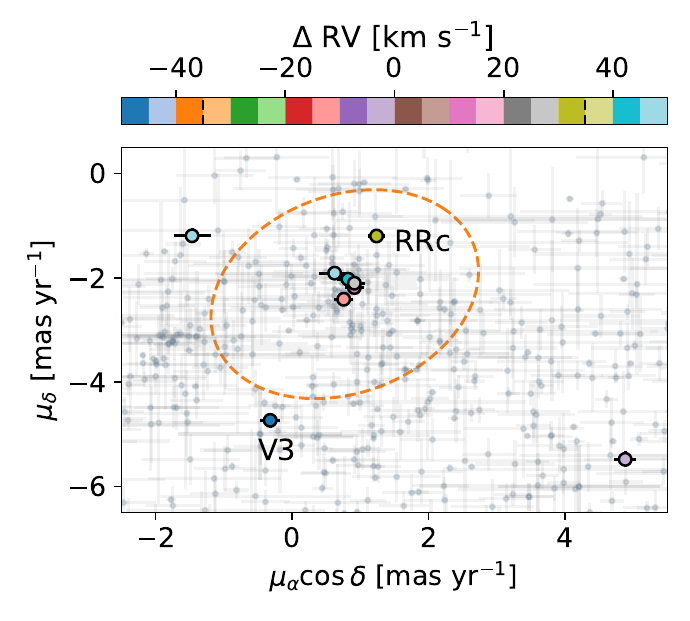}
\includegraphics[width=0.35\textwidth]{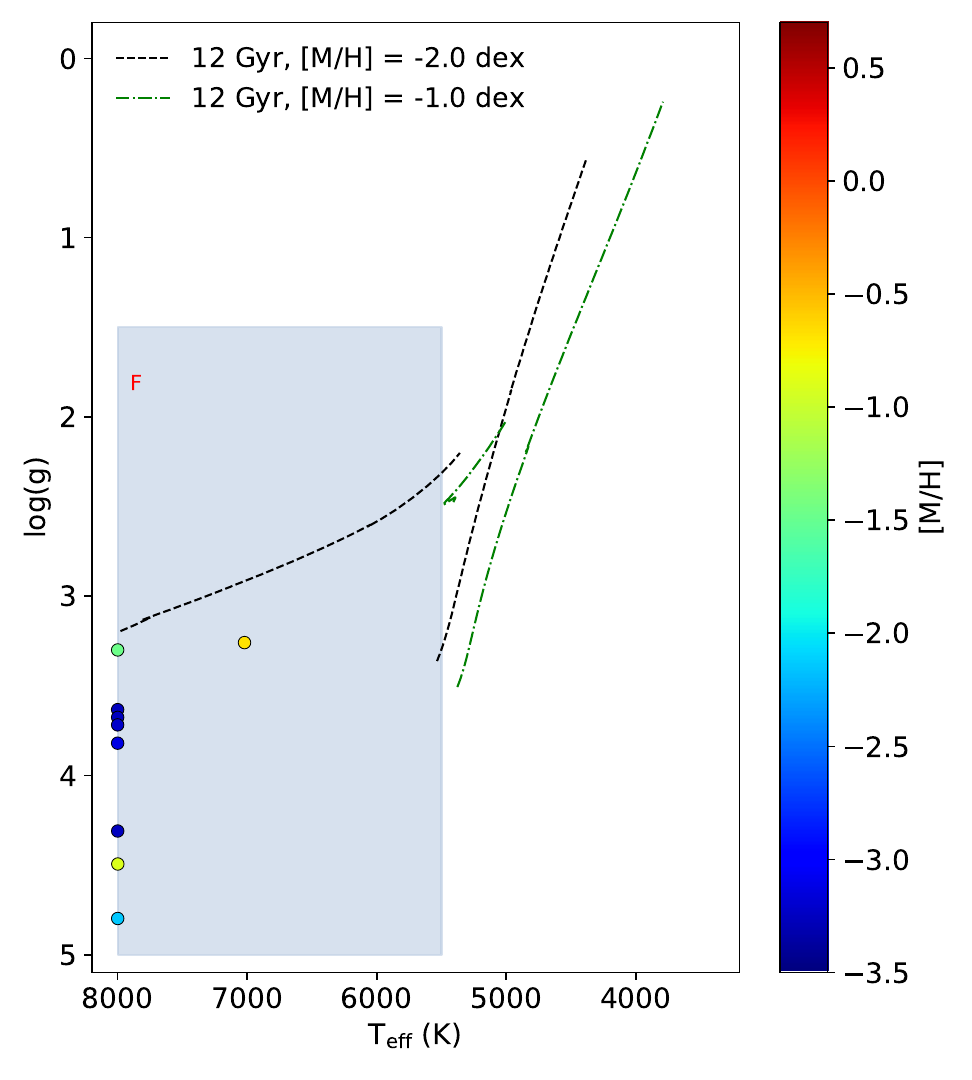}
\caption{Properties of the BHB candidates. Left panel: Proper motion distribution of the BHB candidates in our sample. Each star is colored based on the difference in RV with respect to the mean RV of the cluster. The vertical dashed lines in the colorbar mark the limits of the RV cut imposed to select extra-tidal stars and the dashed orange ellipse the corresponding proper motion cut. Right panel: Kiel diagram showing the evolutionary tracks for horizontal branch stars and the atmospheric parameters recovered for the BHB candidates. Most of them are in the border of the grid, meaning that they probably have T$_{\rm eff}\geq$ 8\,000 K.}\label{fig:bhbs}
\end{figure}

In Table~\ref{tab:bhb}, we listed these BHB and RR Lyrae stars, including their derived RVs, S/N ratio, angular separation from the cluster center, heliocentric distance (derived as described in Section~\ref{sim}) and additional remarks. Metallicities are not reported as most of them are at the edge of the grid to derive the atmospheric parameters (see right panel in Fig.~\ref{fig:bhbs}). The left panel of Fig.~\ref{fig:bhbs} shows the proper motion distribution and RV difference with respect to the cluster mean RV, showing that some BHB stars passed the proper motion cut but the $\Delta$RV is larger than the selection done in this work.

\begin{table*}[h!]
    \centering
    \caption{RV measurements for BHB candidates and RR Lyrae stars, selected based on their position in the CMD}
    \begin{tabular}{ccccccc}
   \hline
 \textit{Gaia} DR3 source\_id  & RV            & $\sigma_{\rm RV}$ & SN & r        & Distance & Remarks \\
                               & (km s$^{-1}$) & (km s$^{-1}$)     &    & (arcmin)  &  (kpc)  &     \\
\hline
2409617596703174016 & -191.0 & 2.6 & 25 & 3.5  & 23.6 & Compatible with the cluster in proper motion and RV \\  
2409615878716218112 & -361.0 & 0.8 & 17 & 5.1  & 30.3 & V3, RRc field star, incompatible proper motion and RV\\ 
2409577125226608128 & -165.5 & 1.2 & 26 & 32.1 & 23.9 & Extra-tidal star candidate \\  
2407021615390864128 & -123.5 & 0.8 & 19 & 38.9 & 28.1 & Incompatible proper motion and RV \\  	
2410199994268808064 & -142.0 & 0.2 & 25 & 61.7 & 25.1 & Extra-tidal star candidate, RRc star\\  		
2410227138462276736 & -136.5 & 1.4 & 16 & 80.7 & 24.8 & Incompatible RV \\  
2410215868467790720 & -130.0 & 3.6 & 12 & 85.0 & 26.0 & Incompatible RV  \\  
2410257688564460928 & -151.1 & 3.4 & 17 & 95.1 & 25.6 & Extra-tidal star candidate \\ 
2406419254817086976 & -181.0 & 11.5 & 20 & 102.7 & 32.7 & Incompatible proper motion \\  	
\hline
    \end{tabular}
    \label{tab:bhb}
\end{table*}
\newpage

\section{Cluster star and extra-tidal star candidates} \label{app:extra_tidal}

Stars inside and outside the tidal radius of NGC 7492, selected based on \textit{Gaia} proper motions and CMD, and GIRAFFE RVs are listed in Table \ref{tab:extra}, including [M/H] only for stars with spectra having S/N $\geq$ 10.

\begin{table*}[h!]
    \centering
    \caption{FLAMES radial velocities and [M/H] for stars associated with the cluster.}
    \begin{tabular}{ccccccc}
   \hline
   \textit{Gaia} DR3 source\_id  & RV  & $\sigma_{\rm RV}$ & [M/H] & $\sigma_{\rm [M/H]}$ & SN & r\\
                    & (km s$^{-1}$) & (km s$^{-1}$) &     & &   & (arcmin) \\
   \hline
      \hline
2409630485899742976 & --178.2 & 0.6 & --1.97 & 0.07 &  18 & 0.4  \\
2409618700509475968 & --177.0 & 0.1 & --2.03 & 0.02 &  47 & 0.4 \\
2409630451540559616 & --174.9 & 0.1 & --1.87 & 0.01 &  88 & 0.6 \\
2409618730576256128 & --175.6 & 1.0 & --1.79 & 0.06 &  17 & 0.7 \\
2409618734869772160 & --178.3 & 0.1 & --1.75 & 0.01 &  80 & 1.1 \\
2409619108531707136 & --173.8 & 0.4 & --2.07 & 0.04 &  24 & 1.4 \\
2409630863856865152 & --172.5 & 0.5 & --1.83 & 0.05 &  20 & 1.7 \\
2409619039812224896 & --173.8 & 0.1 & --1.85 & 0.01 &  66 & 2.0 \\
2409619039812620416 & --174.2 & 0.2 & --1.96 & 0.02 &  31 & 2.4 \\
2409618455696652928 & --176.2 & 0.1 & --1.89 & 0.01 &  74 & 2.4 \\
2409618352617420928 & --172.8 & 0.5 & --2.46 & 0.08 &  19 & 2.5 \\
2409631134440174976 & --172.5 & 0.4 & --1.84 & 0.04 &  22 & 2.6 \\
2409618837949222656 & --176.0 & 0.1 & --1.96 & 0.01 &  93 & 2.9 \\
2409631443677809408 & --174.0 & 0.4 & --1.89 & 0.04 &  19 & 4.1 \\
2409632233951827072 & --175.9 & 0.4 & --2.38 & 0.07 &  16 & 5.4 \\
2409631688490608000 & --174.3 & 0.2 & --1.81 & 0.02 &  34 & 6.4 \\
2409631787275215488 & --168.5 & 2.8 &  --1.95 & 0.59 &  11 & 6.7 \\
2409614195089038720 & --174.2 & 0.3 & --1.91 & 0.03 &  27 & 6.7 \\
2410006961258396288 & --175.3 & 0.1 & --1.75 & 0.01 &  94 & 8.3 \\
2409614092009797888 & --174.9 & 0.2 & --2.13 & 0.02 &  30 & 8.3 \\
\hline
2409605467715384064 & --172.0 & 0.5 & --1.46 & 0.06 &  21 & 14.6 \\
2409995794343427328 & --170.0 & 0.7 & --2.78 & 0.12 &  17 & 15.8 \\
2406610707279564160 & --183.7 & 0.6 & --2.26 & 0.08 &  16 & 16.3 \\
2409991293217682176 & --182.2 & 0.8 & --0.94 & 0.14 &  17 & 16.6 \\
2409677176489588608 & --175.5 & 1.4 & --1.98 & 0.10 &  15 & 28.3 \\
2410006067905281024 & --166.6 & 0.3 & --2.11 & 0.04 &  25 & 28.9 \\
2409599557840380032 & --180.3 & 2.8 & --2.18 & 0.78 &  13 & 30.2 \\
2409648524762962432 & --183.5 & 0.1 & --1.84 & 0.01 &  100 & 30.7 \\
2406574870072342528 & --177.6 & 0.1 & --1.72 & 0.01 &  119 & 31.5 \\
2409577395809021952 & --179.1 & 0.9 & --1.86 & 0.09 &  16 & 31.9 \\
2409598183450835584 & --167.8 & 2.0 & --0.77 & 0.16 &  12 & 34.3 \\
2409579805286184448 & --181.7 & 0.1 & --1.60 & 0.01 & 169 & 36.6 \\
2410043352516444288 & --170.4 & 0.1 & --1.78 & 0.02 &  37 & 38.0 \\
2409573650597520512 & --188.5 & 2.0 & --1.33 & 0.30 &  10 & 40.7 \\
2410203602041372032 & --153.7 & 3.2 &        &      &   9 & 70.5  \\
2406438363126723968 & --189.1 & 0.1 & --1.42 & 0.01 &  116 & 72.9 \\
2406405510921642368 & --196.0 & 4.2 & --1.57 & 0.26 &   12 & 107.2 \\
\hline
    \end{tabular}
    \label{tab:extra}
\end{table*}

\section{Mean RV and metallicity of NGC~7492} \label{app:lit_values}

NGC~7492 has limited measurements of its mean RV and metallicity. In this work, we have collected 20 stars inside the tidal radius of the cluster, having compatible proper motions and that are located along the colour-magnitude diagram of the cluster. We derive the median [M/H] from the full spectrum fitting with FERRE, along with the robust MAD (1.48 times MAD) and listed the most recent values for the [Fe/H] of the cluster in Table~\ref{tab:feh_literature}. Our measurement is robust and based on a significantly larger number of stars compared to previous works. However, the metallicities derived with FERRE tend to be underestimated, and that could explain why our measurement is the most metal-poor one.

\begin{table*}[h!]
    \centering
    \caption{Literature values for the metallicity of NGC~7492, and the value derived in this work.}
    \begin{tabular}{cccc}
   \hline
   [Fe/H] or [M/H]& Reference & Method & Nstars\\
                       &           &        &       \\
   \hline
      \hline      
 --1.89$\pm$ 0.11  & This work & Full spectrum fitting & 19 \\
 --1.64 $\pm$0.13 & \citet{FigueraJaimes13} & Fourier light-curve decomposition of the RR Lyrae stars & 1 \\
 --1.83 $\pm$0.07 & \citet{Cohen05} & Fe I lines in bright red giants in this cluster & 4\\
 --1.79 $\pm$0.08 & \citet{Cohen05} & Fe II lines in bright red giants in this cluster & 4\\
 \hline
    \end{tabular}
    \label{tab:feh_literature}
\end{table*}

For the RVs, we derived a median value of $-$174.9$\pm$0.5 km s$^{-1}$ in excellent agreement with previous measurements, within the errors. We have derived the RV dispersion as the quadratic difference between the observed RV dispersion and the median of the individual RV errors, and its uncertainty as $\Delta \sigma_{\rm RV} \approx \sigma_{\rm RV}/\sqrt{2(N-1)}$. The measured RV dispersion is 1.6$\pm$0.3 km s$^{-1}$, similar to previous works in the literature, as reported in Table~\ref{tab:rv_literature}.

\begin{table*}[h!]
    \centering
    \caption{Literature values for the RV and RV dispersion, $\sigma_{\rm RV}$ of NGC~7492, and the values derived in this work.}
    \begin{tabular}{cccc}
   \hline
   RV           & $\sigma_{\rm RV}$ & Nstars & Reference \\
   (km s$^{-1}$)  & (km s$^{-1}$)     &        &       \\
   \hline
      \hline     
 $-$174.9$\pm$0.5 & 1.6$\pm$0.3 & 20 & This work \\
 $-$175.4$\pm$0.4 & 1.7$\pm$0.5 & 21 & \citet{Geha26a, Geha26b}  \\    
 $-$176.7$\pm$0.2 &             & 29 & \citet{BaumgardtHilker18} \\
 $-$176.9$\pm$0.6$^{a}$ & 1.2$\pm$1.0 & 4 & \citet{Cohen05} \\
 $-$214.2$\pm$11.5 & & 11 & \citet{Rutledge97}\\
 \hline
    \end{tabular}
\vspace{1mm}

\begin{minipage}{0.6\textwidth}
\footnotesize
$^{a}$ Error in the mean computed from the individual stellar RV measurements reported in \citet{Cohen05}.
\end{minipage}

\label{tab:rv_literature}
\end{table*}

\section{Extra-tidal stars from \cite{Chenstreams}}\label{app:chen_stream}

Recently, \cite{Chenstreams} presented a sample of stars associated with different GCs, aiming to derive their mass loss rates. In the sample, NGC 7492 is included, obtaining that among the low-mass cluster, it has a high $\dot{M} \sim$ 9.0 M$_{\odot}$ yr$^{-1}$. Given its relatively large half-mass radius r$_{\rm h}\sim$10.6 pc, NGC 7492 appears to be close to complete tidal dissolution.

\begin{figure}[h!]
\centering
\includegraphics[width=0.5\textwidth]{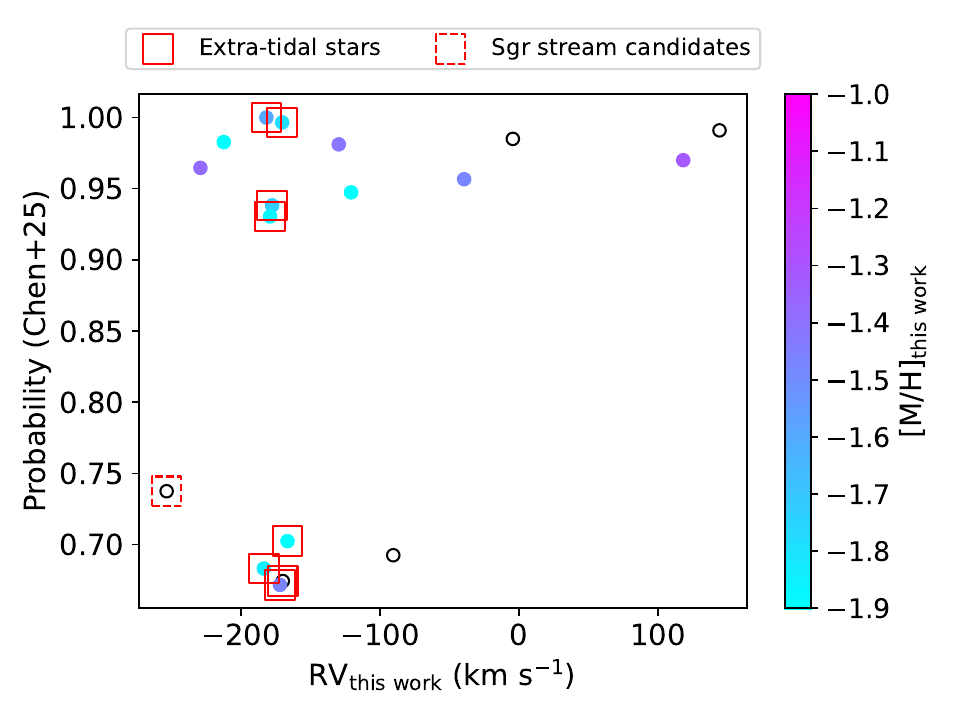}
\caption{RV and metallicity from FLAMES spectra, as derived in this work, and the probability of being associated with the cluster, for the stars in common with \cite{Chenstreams}.} \label{fig:RV_prob}
\end{figure}

We cross-matched the sample of 96 stars recovered in that work with our sample of stars with FLAMES spectra, getting 18 stars in common. Figure~\ref{fig:RV_prob} shows the RV and probability from \cite{Chenstreams}, each point colored by their spectroscopic metallicity for those spectra with S/N $\geq$ 10. For the low S/N spectra, the stars are shown as unfilled black circles. This plot shows that stars with probabilities close to 1.0 have RV values consistent with the GC population as well as consistent with the field population, constituting outliers for the determination of the mass loss rate. 

All the 18 stars in common with \cite{Chenstreams} passed our proper motion cut, however eight of them are already identified in our work as associated to the cluster (marked with unfilled red boxes), only one star is associated to the Sgr stream (marked with an unfilled, dashed red box) and the other nine stars are most likely field stars based on their RVs and/or distance to the isochrone of the cluster. 

\section{Rotation matrix} \label{app:rotation}

We define a local tangent-plane coordinate system centered on NGC 7492, $x = (\alpha - \alpha_0) \cos{\delta_0}$ and $y = (\delta - \delta_0)$, where ($\alpha_0, \delta_0)$~=~(347.112, --15.611) degrees is the cluster center. To align the coordinate system with the morphological axis of the stream, we rotated the tangent-plane coordinates by an angle $\phi$ = 49.416 degrees, measured counter-clockwise from $+x$). The stream-aligned coordinates ($\phi_1$,~$\phi_2$) are then given by
\begin{center}
\begin{equation}
\begin{pmatrix}
\phi_1 \\[3pt]
\phi_2
\end{pmatrix}
=
\begin{pmatrix}
 0.65056223  & 0.75945289 \\
-0.75945289  & 0.65056223 
\end{pmatrix}
\begin{pmatrix}
x \\[3pt]
y
\end{pmatrix}
\end{equation}
\end{center}

In this convention, $\phi_1$ measures the perpendicular displacement from the stream, and 
$\phi_2$ increases along the stream.

\section{Radial velocity gradient}\label{app:rv_gradient}

\begin{figure}[h!]
\centering
\includegraphics[width=0.5\textwidth]{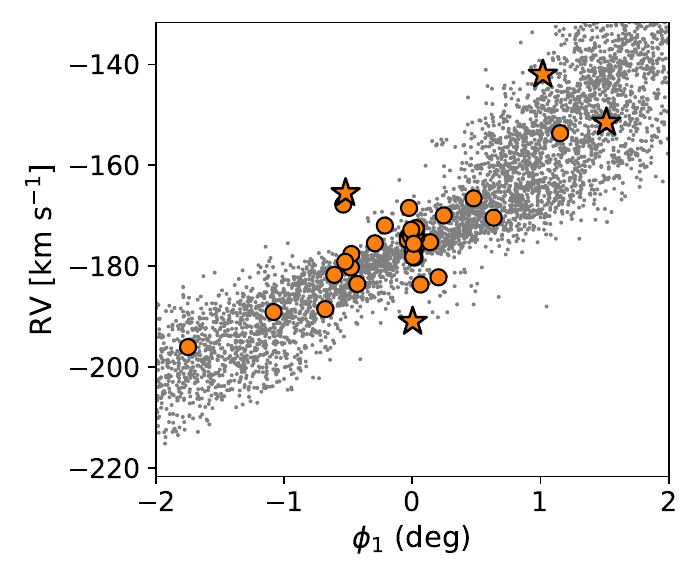}
\caption{RV as a function of $\phi_1$ for the simulated particles (grey dots), the cluster and extra-tidal stars (orange circles) and the extra-tidal BHB stars (orange stars). A prominent RV gradient is recovered.} \label{fig:rv_gradient}
\end{figure}

In Section 5.1, the distribution of simulated particles and our cluster and extra-tidal stars have been compared, particularly on their spatial distribution, radial velocity and distance gradients. From the simulations, a RV gradient going from $\Delta$RV~=~$-$30 to +30 km s$^{-1}$ in the range of -2.0 $\leq \phi_1 \leq$ 2 deg was recovered. From the observations, the RV gradient was not as evident as shown in Figure~\ref{fig:df_delta}. To better visualize the RV gradient, Figure~\ref{fig:rv_gradient} shows the RV as a function of $\phi_1$ for the \cite{Chen25} simulations as the selected stars from our observations (BHB stars are shown with star symbols). Even though the BHBs tend to deviate the most from the predictions of the simulation, the rest of the stars nicely follow the gradient. From the simulation, we were able to fit a gradient consistent with a variation of 16.7 km s$^{-1}$ deg$^{-1}$, similar to what is obtained using the observations (excluding BHBs), of 12.1 km s$^{-1}$ deg$^{-1}$, with respect to the mean RV of the cluster.

\end{appendix}
\end{document}